\documentclass[twocolumn,showpacs,preprintnumbers,amsmath,amssymb,natbib,noshowpacs]{revtex4-1}

\usepackage{graphicx}
\usepackage{dcolumn}
\usepackage{bm,bbm}
\usepackage{color}
\usepackage[colorlinks=true]{hyperref}
\usepackage{amssymb}
\usepackage{amsmath,empheq}
\DeclareMathOperator{\sech}{sech}

\newcommand{\id}{\mathbbm{1}}

\makeatletter
\def\l@subsection#1#2{}
\def\l@subsubsection#1#2{}
\makeatother

\begin{document}

\preprint{OU-HET-919}

\title{Edge-of-edge States}

\author{Koji Hashimoto}
\email{koji@phys.sci.osaka-u.ac.jp}
 \affiliation{Department of Physics, Osaka University,
 Toyonaka, Osaka 560-0043, Japan.}
\author{Taro Kimura}%
 \email{taro.kimura@keio.jp}
\affiliation{%
Department of Physics, Keio University, Kanagawa
 223-8521, Japan.}
\author{Xi Wu}
\email{wuxi@het.phys.sci.osaka-u.ac.jp}
\affiliation{Department of Physics, Osaka University,
 Toyonaka, Osaka 560-0043, Japan.}
%


\begin{abstract}
We study an exotic state which is localized only at an intersection of edges of a 
topological material.
This ``edge-of-edge'' state is shown to exist generically.
We construct
explicitly generic edge-of-edge states in 5-dimensional Weyl semimetals
and their dimensional reductions, such as 4-dimensional topological insulators of class A and 
3-dimensional chiral topological insulators of class AIII.
The existence of the edge-of-edge state is due to a topological charge of the edge states.
The notion of the Berry connection is generalized to include the space of all possible boundary conditions,
where Chern-Simons forms are shown to be nontrivial.
\end{abstract}

\maketitle



\section{Introduction and summary}

Due to the bulk-edge correspondence \cite{Jackiw:1975fn,hatsugai1993chern,Wen:2004ym}
for topological phases \cite{hasan2010colloquium,qi2011topological}, 
edge states are used as a characterization of the nontrivial topology of materials.
The theoretical idea has led to a tremendous success in condensed matter physics,
and various topological materials were discovered experimentally. 

In this paper we
introduce the notion of ``edge-of-edge states'' which is a generalization of the edge states,
and study their existence and implications. In general, materials are surrounded
by many boundaries, and therefore, the boundaries intersect with each other. If we call the
original single boundary as a codimension-1 surface, then the intersection of two
distinct boundaries define a codimension-2 surface. The question is --- are there any
localized states on the intersection? The answer we find is yes, and we call them ``edge-of-edge states.''

The intuition comes from an analogy to D-branes in string theory.
K-theories have been used for the classification of the D-branes \cite{Witten:1998cd}, while they were also used
for the classification of the topological phases~\cite{Schnyder:2008tya,Kitaev:2009mg}. 
In fact, a D-brane on which a gapless fermion lives can be regarded as a surface defect in
a higher-dimensional unstable D-brane. Now, in string theory, when two D-branes intersect,
there generically appear localized modes at the intersection,
when a certain set of conditions for the species of the intersecting D-branes is met.
Therefore, naturally, we may expect such a localized state --- the ``edge-of-edge state'' ---
for topological materials. Clarifying the existence condition of such a state provides 
a new characterization of topological materials.

%

Of course, when the two boundaries are of the same type, there should not exist such an
edge-of-edge state, because the intersection can always be smoothed out. Therefore 
the two boundaries have to have different boundary conditions. Various boundary conditions
can be introduced in topological materials experimentally, but here, we concentrate
on all possible theoretical boundary conditions at the continuum limit.

For the 3D Weyl semimetals, which were recently observed in experiments~\cite{Xu:2015Science,Huang:2015NC,Weng2015:PRX} through theoretical predictions~\cite{Murakami:2007bx,Murakami:2007NJP,Wan:2011PRB,Yang:2011PRB,Burkov2011:PRL,Xu:2011dn,Burkov2011:PRB}, 
generic boundary conditions in the continuum theory were classified in our previous work \cite{Hashimoto:2016kxm}
(see \cite{isaev2011bulk,Enaldiev:2015JETP} for generic boundary conditions 
for topological insulators). The 3D Weyl semimetal has a simple Hamiltonian of $2\times 2$, but it will
turn out that the structure is not large enough to support the existence of the edge-of-edge state.

\begin{figure*}[t]
	\begin{center}
	 \includegraphics[width=50em]{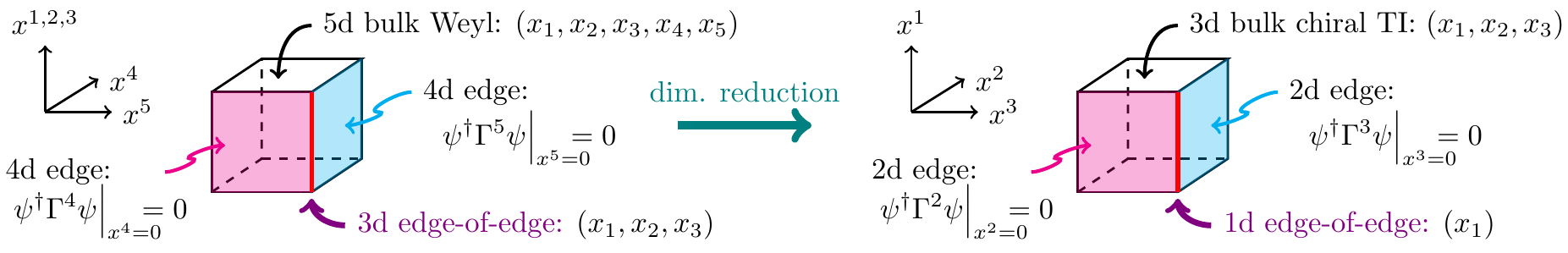}
	 \caption{A schematic picture of the edge-of-edge state and the dimensional reduction from 5d Weyl semimetal to 3d chiral topological insulator (class AIII). There are two boundaries which intersect each other.}
	 \label{two-b}
          \end{center}
\end{figure*}

Recently, in the context of lattice gauge theories, intersection of two distinct boundaries
in $1+5$ dimensional spacetime was studied \cite{Fukaya:2016dcl,Fukaya:2016ofi} for realizing a regularization of
chiral gauge field theories. The Dirac operator in that dimension is an $8\times 8$ 
matrix, which was shown to accommodate localized chiral mode at the intersection,
under a particular set of the boundary conditions at the boundaries. Encouraged
by this example, we are led to the present study which clarifies the least spatial dimensions and the size of the Hamiltonian to accommodate any possible edge-of-edge
state.

In this paper we consider a minimal case which allows the edge-of-edge states: Hamiltonians of the size of $4\times 4$. 
The simplest realization is
a 5-dimensional Weyl fermion \cite{PhysRevB.94.041105} and its dimensional reductions, in particular, a 3D chiral topological insulator of class AIII.
The Hamiltonian of the 5D
Weyl semimetal is given by a generalization of that of the 3D Weyl semimetals,
${\cal H} = \sum_{M=1}^5 \Gamma^M p_M$, 
where $\Gamma^M$ is the $4\times 4$ Gamma matrix. 
Its dimensional reduction with a mass $p_4=m$ and $p_5=0$ leads to
the 3D chiral topological insulator of class AIII. (Two more concrete examples in 3D topological insulator were studied in \cite{Sen:2012aa} and \cite{Sen2014}, in which the boundary parameters appear as potential barriers.)
For this size of the Hamiltonians, typically there could appear two edge
states for a single boundary. And at the intersection of the two boundaries,
the edge-of-edge state can appear.

To derive the edge-of-edge states for the 3D chiral topological insulators of class 
AIII, it is instructive to work first for the 5D Weyl semimetals. So, in this paper 
first we work in the 5D case, and then make a dimensional reduction to the three
dimensions.

Our findings for 
the 3D chiral topological insulators of class AIII 
(and for the continuum 5D Weyl semimetals)
in this paper are summarized below.
\begin{itemize}
\item
Generic boundary conditions are dictated
by a $U(2)$ parameter.
\item
The edge-of-edge states can exist generically.
(See Fig.~\ref{two-b}.)
\item
Existence condition of the edge-of-edge state is derived.
\item
The edge-of-edge states is gapless, while the edge states could be gapped.
\item
The edge states have topological charges
characterized by Chen-Simons integrals.
\end{itemize}

The organization of this paper is as follows. In section \ref{sec:3D}, we review the
generic boundary conditions of the 3D Weyl semimetals, following \cite{Hashimoto:2016kxm}.
From section \ref{sec:5D}, we study the $4\times 4$ Hamiltonians: 
the simplest 5D Weyl semimetal is used for the analyses to be transparent.
In section \ref{sec:5D}, we 
obtain generic boundary conditions of the 5D Weyl semimetal
and edge states with their dispersions. In section \ref{sec:eoe}, we discover the
edge-of-edge states and obtain the existence condition and the generic dispersion
relation of them. We study the mechanism of the edge of edge states.
In section \ref{sec:AIII}, we study the dimensional reduction to the 3D chiral topological
insulator of class AIII, and see that all our arguments about the edge-of-edge states
apply similarly. In section \ref{sec:top}, we analyze the
topological charges of the edge states. 
The final section is for various discussions, and the appendix for detailed
calculations.


\section{Review: boundary condition in 3D}
\label{sec:3D}

Let us briefly summarize generic boundary conditions of 3D Weyl semimetals in
the continuum limit, following our previous paper \cite{Hashimoto:2016kxm}.
It guides us to find generic boundary conditions of 5D Weyl semimetals in
the next section.

The 3D Weyl semimetal Hamiltonian near the tip of the Weyl cone is
\begin{align}
{\cal H} = p_i \sigma_i \, .
\end{align}
and the Hamiltonian eigen equation is
\begin{align}
p_i \sigma_i \psi = \epsilon \psi \, .
\label{Hamil3D}
\end{align}
Our metric convention is chosen as $\eta_{\mu\nu}=\mbox{diag}(+,-,-,-)_{\mu\nu}$. 
$\sigma^\mu=(\id_2, \sigma_1,\sigma_2,\sigma_3)$.

The total action is
\begin{align}
S = \int_{x^3\geq 0} \!\!\!\!d^4x \; 
\frac{i}{2} \psi^\dagger \sigma^\mu (\overrightarrow{\partial}_\mu 
 - \overleftarrow{\partial}_\mu)
\psi
 + \frac12 \int_{x^3=0} \! d^3x \;
\psi^\dagger N \psi \, .
\end{align}
The first term is the Weyl Lagrangian. 
The second integral is with a Hermitian matrix $N$.
The boundary condition follows from this Lagrangian as
\begin{align}
(M+\id_2) \psi\Big|_{x^3=0} = 0,
\end{align}
with $N =  i \sigma_3 M$.

%

The Hermiticity $N^\dagger = N$ and the vanishing determinant condition
$\det (M+1)=0$ leads to a generic solution
\begin{align}
	M=A_1\sigma_1+A_2\sigma_2+iB_3\sigma_3,
	\\
\mbox{with} \quad	A_1^2+A_2^2-B_3^2=1.
\end{align}
We can choose
\begin{empheq}[left=\empheqlbrace]{align}
		\nonumber &A_1=\cos\theta\cosh\chi\\  
		&A_2=\sin\theta\cosh\chi \\
		\nonumber &B_3=\sinh\chi
\end{empheq}
for parametrizing the matrix. 
Defining $\cos{\theta'}=\sech{\chi}$ and $\sin{\theta'}=\tanh{\chi}$ and changing variables: 
\begin{align*}
	\theta'&=\theta_++\theta_-
	\\
	\theta&=\theta_+-\theta_-
\end{align*}
the boundary condition becomes
\begin{align}
	\left(\begin{array}{cc}
	e^{i\theta'} & e^{-i\theta}  
	\\ 
	e^{i\theta} & e^{-i\theta'} 
	\end{array}\right)
	\psi\Big|_{x^3=0}=0.
\end{align}
Noting a relation 
\begin{align}
\left(\begin{array}{cc}
	e^{i\theta'} & e^{-i\theta}  
	\\ 
	e^{i\theta} & e^{-i\theta'} 
	\end{array}\right)
	=
	\left(\begin{array}{c}
	e^{i\theta'} \\
	e^{i\theta} 
	\end{array}\right)
	\left(\begin{array}{cc}1 & e^{-2i\theta_+} 
	\end{array}\right),
\end{align}
the boundary condition is recast to the following simple form
\begin{align}\label{bc'}
	\left(\begin{array}{cc}1 & e^{-2i\theta_+} 
	\end{array}\right)
	\psi\Big|_{x^3=0}=0.
\end{align}

The dispersion relation of the edge mode is
\begin{align}\label{EDR}
	\epsilon=-p_1\cos{2\theta_+}-p_2\sin{2\theta_+}.
\end{align}
And the general edge mode wave function is 
\begin{align}
\label{edgegeneral}
	\psi(x^3)&=\sqrt{\alpha}~\text{exp}(-\alpha x^3)
	\left(\begin{array}{c} e^{-2i\theta_+} \\ -1
	\end{array}\right),
	\\ \alpha&=p_1\sin{2\theta_+}-p_2\cos{2\theta_+}. 
	\label{alpha12}
\end{align}
The edge mode exists only in a limited region of the momentum space $\alpha(p)>0$. 

So, in summary, the generic boundary condition \eqref{bc'} is dictated by a single real $U(1)$ parameter $\theta_+
\in S^1$.
In the following, we will find that the 5D generalization is dictated by a $U(2)$ parameter.



\section{5D Weyl semimetals}
\label{sec:5D}

\subsection{Hamiltonian, Lagrangian and surface term}

The Weyl fermion in 1+5 spacetime dimensions has the Hamiltonian
\begin{align}
{\cal H} = \sum_{M=1}^5 \Gamma^M p_M
\label{Hamil5}
\end{align}
as in the same manner as the standard Weyl semimetal Hamiltonian 
${\cal H}=p_1 \sigma_1 + p_2 \sigma_2 + p_3 \sigma_3$ in 1+3 spacetime 
dimensions. Here $\Gamma^M$ ($M=1,\cdots,5$) is the $4\times 4$ Gamma
matrix satisfying 
the 5-dimensional Euclidean Clifford algebra
\begin{align}
\{\Gamma^M,\Gamma^N\} = 2\delta^{MN} \quad (M,N=1,2,3,4,5)
\label{G5}
\end{align}

Upon a dimensional reduction to 1+4 dimensions by replacing $p_5$ by
a constant $m$, the system reduces to the class A topological insulator
in 4 dimensions with the Hamiltonian
\begin{align}
{\cal H} = p_i \Gamma^i + m\Gamma^5 \, .
\end{align}

To derive consistent boundary conditions, we go to a Lagrangian
formulation. 
The bulk Lagrangian is written in the same manner as the 1+3-dimensional case.
Now with the gamma matrices in 4+1 dimensions,
\begin{align}
{\cal L} = - \psi^\dagger i \gamma^0
(\gamma^\mu \partial_\mu -i\partial_5) \psi
\end{align}
with $\bar{\psi}\equiv \psi^\dagger i \gamma^0$.
Here $\mu=0,1,2,3,4$ and the $4\times 4$ 
gamma matrices are a representation of the Clifford algebra 
$\{\gamma^\mu,\gamma^\nu\}=2\eta^{\mu\nu}$ ($\mu,\nu=0,1,2,3,4$). 
Note that the Gamma matrices $\gamma^\mu$ are a part of $8\times 8$ Gamma
matrices in $1+5$ dimensions. The Dirac equation is
\begin{align}
(\gamma^\mu \partial_\mu -i\partial_5) \psi
=0
\end{align}
which can be rewritten as
\begin{align}
\left[i\partial_0 -i\gamma^0(\gamma^i \partial_i -i\partial_5) \right]\psi
=0
\end{align}
where $i=1,2,3,4$.
So the Hamiltonian is $i\partial_0 = {\cal H}$, 
\begin{align}
{\cal H} \equiv -\gamma^0\gamma^ip_i  + i \gamma^0 p_5 \, .
\label{eq:hamil}
\end{align}
We have used $p_i = -i \partial_i$ and $p_5 = -i\partial_5$. 
If we use a redefined Gamma matrices
\begin{align}
\Gamma^5 \equiv i\gamma^0, 
\quad
\Gamma^i \equiv -\gamma^0\gamma^i
\end{align}
then they satisfy \eqref{G5}.
And the Hamiltonian is conveniently written as \eqref{Hamil5}.

The boundary condition is imposed at $x^5=0$,
\begin{align}
A\psi = 0 \, .
\end{align}
Again defining $A = M + \id_4$, we have
\begin{align}
M\psi = - \psi \, .
\label{eq:Mp}
\end{align}
One of the eigen value of $A$ is vanishing.
The Hamiltonian self-conjugacy condition leads to
\begin{align}
M^\dagger \Gamma^5 + \Gamma^5 M = 0 \, .
\label{eq:M4}
\end{align}

For the Lagrangian formalism, we have an action with a surface term
\begin{align}
S =& -\int d^6x \; 
 \bar{\psi}\left(\frac12 
 (\gamma^\mu \overrightarrow{\partial}_\mu -i\overrightarrow{\partial}_5)
 -\frac12 (\gamma^\mu \overleftarrow{\partial}_\mu-i\overleftarrow{\partial}_5)\right) \psi
 \nonumber \\
 &
 + \frac12 \int_{x^5=0} \! d^5x \;
\psi^\dagger N \psi \, .
\label{eq:2dL}
\end{align}
Here $N$ is a Hermitian $4 \times 4$ matrix.
Following the same logic as in the 1+3-dimensional case, 
we arrive at the boundary condition derived from this action as
\begin{align}
 (\id_4 -\gamma^0 N) \psi(x^5=0) = 0 \, .
 \label{bcN4}
\end{align}
With a definition $N =  \gamma^0 M$, we can reproduce
the boundary condition (\ref{eq:Mp}). By the Hermiticity of $N$, the matrix 
$M$ needs to satisfy (\ref{eq:M4}).


\subsection{Generic boundary conditions}

The boundary condition (\ref{bcN4}) is
\begin{align}
(\id_4 + i \Gamma^5 N)\psi\Big|_{x^5=0} = 0.
\label{bou5}
\end{align}
We want to know what is the generic solution $\psi$ of this equation.
See also Appendix \ref{sec:BC} for the boundary condition imposed to other boundaries.
Suppose there are two solutions, $\psi_1$ and $\psi_2$. Then we can show
for any $\psi_1$ and $\psi_2$
\begin{align}
\psi_1^\dagger \Gamma^5\psi_2 = 0 \, .
\label{152}
\end{align}
The reason is simple: using \eqref{bou5}, we obtain
\begin{align}
\psi_1^\dagger \Gamma^5\psi_2 
&=\psi_1^\dagger (-iN)\psi_2 
=(iN\psi_1)^\dagger \psi_2 
=(-\Gamma^5 \psi_1)^\dagger \psi_2 \nonumber \\
&=-\psi_1^\dagger \Gamma^5\psi_2 \, ,
\end{align} 
which means \eqref{152}.
In this paper, we use the following representation of the Clifford algebra, 
\begin{subequations}
 \label{eq:Gamma_matrices}
\begin{align}
&\Gamma^i = \left(
\begin{array}{cc}
0& -i\sigma_i \\ i\sigma_i & 0
\end{array}
\right),
\\
&\Gamma^4 = \left(
\begin{array}{cc}
0& \id_2 \\ \id_2 & 0
\end{array}
\right),
\quad
\Gamma^5 = \left(
\begin{array}{cc}
\id_2 & 0 \\ 0 & -\id_2
\end{array}
\right).
\end{align}
\end{subequations}
Then, decomposing $\psi_1 = (\xi_1,\eta_1)^T$ and $\psi_2=(\xi_2,\eta_2)^T$,
\eqref{152} is equivalent to
\begin{align}
\xi_1^\dagger \xi_2 - \eta_1^\dagger \eta_2 = 0 \, .
\end{align}
This equation is satisfied only if
\begin{align}
\eta_1 = U_5 \xi_1 \, , \quad
\eta_2 = U_5 \xi_2 \, , \quad
\end{align}
for an arbitrary $U(2)$ matrix $U_5$. So, we conclude that the 
consistent generic solution of the boundary condition \eqref{bou5} is
\begin{align}
\psi \propto \left(
\begin{array}{c}
\id_2 \\ U_5
\end{array}
\right)\xi
\label{bou55}
\end{align}
for a normalized two-spinor $\xi$.
We remark that it can be reparametrized using $U(2)$ rotation, $\xi \to V \xi$ with $V \in U(2)$.
In other words, the boundary condition is rephrased to
\begin{align}
\left(\id_2 \; -\!U_5^\dagger \right) \psi\Big|_{x^5=0} = 0 \, .
\label{bou55-}
\end{align}
This condition is analogous to the 1+3-dimensional case \eqref{bc'}.
We notice that the previous $e^{2i\theta_+}$ is replaced by the $U(2)$
unitary matrix $-U_5^\dagger$. We have four real parameters to parametrize
the generic boundary condition specified by $U_5$.

The condition \eqref{bou55-} can be written in an alternative manner.
Notice that 
it is equivalent to
\begin{align}
\left(
\begin{array}{cc}
\id_2 &  -U_5^\dagger 
\\
U_5 & -\id_2
\end{array}
\right) \psi\Big|_{x^5=0} = 0 \, .
\label{bou55-+}
\end{align}
In view of the original boundary condition \eqref{bou5}, we find
the relation between the Lagrangian boundary term specified by
the Hermitian matrix $N$ and the boundary condition specified by the
$U(2)$ matrix $U_5$ as
\begin{align}
N = N_5 \equiv \left(
\begin{array}{cc}
0&  iU_5^\dagger 
\\
-iU_5 & 0
\end{array}
\right) \, .
\end{align}
This is just one way to relate \eqref{bou55-} and \eqref{bou5}. There
may be other expressions for $N$ which reproduces \eqref{bou55-},
as in the case of the 3D Weyl semimetals.


\subsection{Edge state}

The bulk Hamiltonian eigen equation for $\psi = (\xi,\eta)^T$ is
\begin{align}
(-i\partial_5-\epsilon)\xi + \left(-i\sigma_i p_i +p_4\right)\eta = 0 
\label{Ham4e1}\\
\left(i\sigma_ip_i +p_4\right)\xi - (-i\partial_5 + \epsilon)\eta = 0
\label{Ham4e2}
\end{align}
with $i=1,2,3$.
The edge state solution to the bulk Hamiltonian eigen equation is
\begin{align}
\psi = \left(
\begin{array}{c}
\xi(p_i,p_4) \\ \eta(p_i,p_4)
\end{array}
\right)\exp[-\alpha_5 x^5] \,, \quad 
\alpha_5 \equiv \sqrt{-\epsilon^2  + p_i^2+p_4^2} \, .
\end{align}
Let us substitute the boundary condition \eqref{bou55}. Then
the equations \eqref{Ham4e1} and \eqref{Ham4e2} are
written as
\begin{align}
\left[(i\alpha_5-\epsilon) + \left(-i\sigma_i p_i +p_4\right)U_5\right]\xi = 0  \, ,
\label{a1}\\
\left[-(i\alpha_5+\epsilon)U_5 + \left(i\sigma_i p_i +p_4\right)\right]\xi = 0  \, .
\label{a2}
\end{align}
Noting that the unitary matrix $U_5$ determining the boundary condition
can be decomposed as
\begin{align}
U_5 = e^{i\theta_5} U_5'
\end{align}
where $U_5'$ is an $SU(2)$ matrix, and this acts as a rotation in the
4-dimensional momentum space,
\begin{align}
\left(-i\sigma_i p_i +p_4\right)U_5'
=-i\sigma_i \tilde{p}_i +\tilde{p}_4
\label{tp}
\end{align}
with
\begin{align}
p_i^2+p_4^2 = \tilde{p}_i^2+\tilde{p}_4^2.
\end{align}
Then the two equations \eqref{a1} and \eqref{a2} are
\begin{align}
\left[e^{-i\theta_5}(i\alpha_5-\epsilon) +\tilde{p}_4-i\sigma_i \tilde{p}_i
\right]\xi = 0  \, ,\\
\left[-e^{i\theta_5}(i\alpha_5+\epsilon)+\tilde{p}_4+i\sigma_i \tilde{p}_i\right]\xi = 0  \, .
\end{align}
Equivalently,
\begin{align}
\left[\alpha_5\sin\theta_5-\epsilon\cos\theta_5 +\tilde{p}_4
\right]\xi = 0  \, ,
\label{equi1}\\
\left[\alpha_5\cos\theta_5+\epsilon\sin\theta_5-\sigma_i \tilde{p}_i\right]\xi = 0  \, .
\label{equi2}
\end{align}
This has a solution only when 
\begin{align}
\alpha_5\sin\theta_5-\epsilon\cos\theta_5 +\tilde{p}_4 = 0  \, ,\\
\det\left[\alpha_5\cos\theta_5+\epsilon\sin\theta_5-\sigma_i \tilde{p}_i\right]= 0  \, .
\end{align}
The second equation implies
\begin{align}
\alpha_5\cos\theta_5+\epsilon\sin\theta_5 = \pm\sqrt{  \tilde{p}_i^2} \, .
\end{align}
So we finally obtain the dispersion relation of the edge state,
\begin{align}
\epsilon & = \tilde{p}_4\cos\theta_5 \pm \sqrt{\tilde{p}_i^2} \sin\theta_5 \, , \\
\alpha_5 & = -\tilde{p}_4\sin\theta_5 \pm \sqrt{\tilde{p}_i^2} \cos\theta_5 \, .
\label{al5}
\end{align}
The normalizability condition is $\alpha_5 > 0$ which constrains the momentum region
for the existence of the edge state.

One may notice the similarity to the 1+3-dimensional case of the standard Weyl semimetals,
\eqref{EDR} and \eqref{alpha12}. In fact, identifying $2\theta_+ =\theta_5 + \pi$
and putting $p_2=p_3=0$ with $U_5'=\id_2$ 
means a consistent reduction from 1+5 dimensions to 1+3 dimensions,
reproducing all the results of the three-dimensional Weyl semimetals.

\section{Edge-of-edge states}
\label{sec:eoe}

\subsection{Introducing another edge}

To realize an intersection of the edges, we need a set of edges. In addition to the generic
edge considered in the previous section at $x^5=0$, let us introduce another one
at $x^4=0$. The construction of the generic edge state at $x^4=0$ is completely parallel to
that of the previous section. Nevertheless, it would be instructive to construct the
generic edge state explicitly, for the later purpose of finding the edge-of-edge state.

We look for a generic solution to the equation at the boundary
\begin{align}
\psi_1^\dagger \Gamma^4\psi_2\Big|_{x^4=0} = 0
\end{align}
which is analogous to \eqref{152}. Its component expression is
\begin{align}
\xi_1^\dagger \eta_2 + \eta_1^\dagger \xi_2 = 0 \, .
\end{align}
A generic solution of this equation is obtained by a rotation in the 4-5 space from the previous one
at $x^5=0$,
\begin{align}
\psi = 
\left(
\begin{array}{c}
\id_2 -U_4 \\
\id_2 +U_4
\end{array}
\right)\chi
(p_i,p_5)
\exp[-\alpha_4 x^4] \,
\label{x4b} 
\end{align}
with an arbitrary two-spinor $\eta$ and a $U(2)$ matrix $U_4$. This $U_4$ 
parametrizes the boundary condition at $x^4=0$.
See Appendix \ref{sec:BC} for more details on the edge state for $x^a = 0$ $(a = 1, 2, 3, 4)$.

The boundary condition at $x^4=0$ can be written also as
\begin{align}
\left(
\begin{array}{cc}
\frac12(U_4^\dagger-U_4) &  \id_2-\frac12(U_4^\dagger + U_4) 
\\
\id_2 +\frac12( U_4^\dagger + U_4) &  -\frac12(U_4^\dagger-U_4) 
\end{array}
\right) \psi\Big|_{x^4=0} = 0 \, .
\end{align}
This is interpreted as the contribution from the boundary term in the Lagrangian,
\begin{align}
(1+ i \Gamma^4 N_4)\psi\Big|_{x^4=0} = 0 \, .
\end{align}
So the boundary term consists of the following Hermitian matrix $N_4$,
\begin{align}
N_4=
\frac{-i}{2} \left(
\begin{array}{cc}
U_4^\dagger-U_4 &  -U_4^\dagger - U_4 
\\
U_4^\dagger + U_4 &  -U_4^\dagger+U_4
\end{array}
\right) \, .
\end{align}


\subsection{Generic edge-of-edge states}

Let us consider both the boundary conditions at the same time. The
expected wave function should be of the form \eqref{x4b} but at the same time
satisfying \eqref{bou55}.
Therefore we demand, with a two-spinor $\chi$ 
(which needs not to be normalized for our purpose here), 
\begin{align}
\psi = 
\left(
\begin{array}{c}
\id_2 -U_4 \\
\id_2 +U_4
\end{array}
\right)\chi
(p_i)
\exp[-\alpha_4 x^4-\alpha_5 x^5] \,, 
\label{x45b}
\end{align}
with
\begin{align}
\left[U_5(\id_2 -U_4)-(\id_2 +U_4)\right]\chi = 0 \, .
\label{consUchi}
\end{align}
The latter is the compatibility condition \eqref{bou55-}.
For that to have a nontrivial solution, we need
\begin{align}
\det\left[\id_2 +U_4-U_5 + U_5 U_4\right]=0 \, .
\label{consUU}
\end{align}
This is a necessary condition for the existence of the edge-of-edge state.
We remark that the condition \eqref{consUchi} is covariant under the rotation
\begin{align}
 (U_4,U_5,\chi) \ \longrightarrow \ (W U_4 W^\dag, W U_5 W^\dag, W\chi) 
 \label{UU_rot}
\end{align}
with $W \in U(2)$.
So there is an equivalence class of the edge-of-edge states related by this $W$.
Later we will see that the edge-of-edge state is parameterized by a five dimensional
manifold which is a fibration of $S^1\times S^1$.


The Hamiltonian eigen equation leads to
\begin{align}
\left[(i\alpha_4-\epsilon) + \left(-i\sigma_i p_i -i\alpha_5\right)U_4\right]\chi = 0  \, ,
\label{a1--}\\
\left[-(i\alpha_4+\epsilon)U_4 + \left(i\sigma_i p_i -i\alpha_5\right)\right]\chi = 0  \, .
\label{a2--}
\end{align}
Together with
\begin{align}
\epsilon^2 = p_i^2-\alpha_4^2-\alpha_5^2 \, ,
\end{align}
we have three equations with three unknowns $(\epsilon, \alpha_4,\alpha_5)$
so they are solved and determine the edge-of-edge state 
dispersion, as follows.


We first solve the existence condition \eqref{consUU} for the boundary conditions. 
We define
\begin{subequations}
\label{eq:U4U5_parametrization}
\begin{align}
 &U_5=e^{i\theta_5}(a_0\id_2+ia_i\sigma^i)=A_0\id_2+A_i\sigma^i \, ,
 \\
 &U_4=e^{i\theta_4}(b_0\id_2+ib_i\sigma^i)=B_0\id_2+B_i\sigma^i \, .
\end{align}
The unitarity of $U_4$ and $U_5$ means
\begin{align}
 a_0^2+a_i^2=b_0^2+b_i^2=1 \, .
\end{align}
\end{subequations}
After some computations, we obtain 
a consistency relation for the dispersion $\epsilon(p)$ 
of the generic edge-of-edge state to satisfy, 
\begin{align}
   A\epsilon^2-2B\epsilon+C=0\, ,
 \label{eq:eoe_disp}
\end{align}
where the coefficients are defined as
\begin{subequations}
 \label{eq:ABC}
 \begin{align}
  &A\equiv 1-\cos^2\theta_4\cos^2\theta_5  \, , 
  \\
  &B \equiv a_ip_i\cos\theta_5\sin^2\theta_4+b_ip_i\cos\theta_4\sin^2\theta_5  \, , 
  \\
  &C\equiv (a_ip_i)^2\sin^2\theta_4+(b_ip_i)^2\sin^2\theta_5- p_i^2\sin^2\theta_5\sin^2\theta_4\, .
 \end{align}
\end{subequations}
See Appendix \ref{sec:der_eoe} for details of the derivation.

If we want to obtain gapless edge-of-edge states,
we need to require $C=0$, which is
\begin{align}
(a_ip_i)^2\sin^2\theta_4+(b_ip_i)^2\sin^2\theta_5- p_i^2\sin^2\theta_5\sin^2\theta_4 = 0 \, . \nonumber
 \end{align} 
It is obvious that this is gapless for the 5D Weyl semimetals,  since 
this is solved by $p_i=0$.
On the other hand, for the dimensionally reduced case, the gapless condition for the edge-of-edge states cannot always be met.
We will discuss the dimensional reduction from 5d Weyl semimetal to 3d chiral topological insulator (class AIII) in Sec.~\ref{sec:AIII}.

In deriving \eqref{eq:eoe_disp}, we need a relation
(see Appendix \ref{sec:der_eoe})
\begin{align}
a_0=b_0=0\, , \quad a_i^2 = b_i^2 = 1\, , \; 
\nonumber \\
a_i b_i = -\cos\theta_4 \cos\theta_5 \, .
\end{align}
This defines the parameter space of the edge-of-edge state.
It is a five dimensional manifold which is a fibration over $S^1\times S^1$
of $(\theta_4,\theta_5)$.


\subsection{Mechanism of edge-of-edge localization}

To clarify how the edge-of-edge states are possible, we present a typical example.
Let us take, as an example, 
\begin{align}
U_4 = \sigma_3 \, , \quad U_5 = \sigma_2 \, ,
\label{UU}
\end{align}
which satisfies \eqref{consUU}. 
Then \eqref{consUchi} is solved by
\begin{align}
\chi = \left(
\begin{array}{c}
1 \\ i 
\end{array}
\right) \, .
\end{align}
Substituting these to \eqref{a1--} and \eqref{a2--}, we obtain
\begin{align}
\left[(i\alpha_4-\epsilon) + \left(-i\sigma_i p_i -i\alpha_5\right)\sigma_3\right]
\left(
\begin{array}{c}
1 \\ i 
\end{array}
\right)
 = 0  \, ,
\\
\left[-(i\alpha_4+\epsilon)\sigma_3 + \left(i\sigma_i p_i -i\alpha_5\right)\right]
\left(
\begin{array}{c}
1 \\ i 
\end{array}
\right)
 = 0  \, .
\end{align}
This is explicitly solved as 
\begin{align}
\epsilon = -p_1\, , \quad
\alpha_4 = p_3 \, ,
\quad
\alpha_5 = p_2 \, .
\label{4352a}
\end{align}
So we obtain an edge-of-edge state with a linear (chiral) dispersion.
The edge-of-edge state exists for $p_3>0$ and $p_2>0$.

Let us consider the meaning of this edge-of-edge state.
Note that this example with the boundary unitary matrices \eqref{UU}
corresponds to
\begin{align}
N_4 = - \Gamma^3 \, , \quad N_5 = - \Gamma^2 \, .
\end{align}
In other words, the boundary conditions are
\begin{align}
(\Gamma^4 - i \Gamma^3) \psi\Bigm|_{x^4=0} = 0 \, , 
\quad
(\Gamma^5 - i \Gamma^2) \psi\Bigm|_{x^5=0} = 0 \, . 
\label{GG45}
\end{align}
In view of the total Hamiltonian is \eqref{Hamil5}, 
these equations mean that the term $p_4\Gamma^4$
could be canceled by $p_3 \Gamma^3$,
and the term $p_5 \Gamma^5$ can be canceled by
$p_2 \Gamma^2$. In fact,  
the boundary condition \eqref{GG45} can be trivially consistent 
with the structure of the Hamiltonian
when
\begin{align}
i p_4 + p_3 = 0 \, , \quad i p_5 + p_2 = 0 \, .
\label{4352}
\end{align}
Then the remaining Hamiltonian is simply ${\cal H}=p_1 \Gamma^1$,
and its dispersion is $E = p_1$. And the condition \eqref{4352} is
nothing but the relation about $\alpha_4$ and $\alpha_5$, \eqref{4352a}.
We remark that the relation between $p_i$ and $\alpha_i$ corresponds to that between Fourier and Laplace transforms with the kernels $e^{i p x}$ and $e^{-\alpha x}$.

Therefore, the mechanism of the edge-of-edge localization is quite simple:
In the Hamiltonian \eqref{Hamil5}, the gamma matrices are paired to 
be annihilated. (In the case above, for the boundary $x^5=0$, $\Gamma^5$
is paired with $\Gamma^2$ and annihilated in the Hamiltonian.)
This annihilation gives a localized wave function at the edge.
When we have two pairs, the localization is independent and we obtain
an edge-of-edge state.

\section{Reduction to 3d chiral topological insulator (class AIII)}
\label{sec:AIII}

In this section, we discuss the dimensionally reduced model, which is a three-dimensional chiral topological insulator (class AIII) towards an experimental realization of the edge-of-edge state.
See, for example, \cite{Wang:2014PRL} for a setup of the class AIII system using ultracold atoms.

\subsection{Edge-of-edge state at $x^{2,3} = 0$}

In order to study the edge-of-edge state in the 3d model, let us first study the edge states of the 5d Weyl fermion \eqref{Hamil5} at the boundaries $x^2 = 0$ and $x^3=0$.
We impose the boundary condition
\begin{align}
 \psi^\dag \Gamma^a \psi\Big|_{x^{a} = 0} = 0
 \quad (a = 2,3)
 \, .
 \label{eq:bc_23}
\end{align}
The edge state and the corresponding spectrum for this boundary condition is discussed in Appendix \ref{sec:BC} in details.
The edge-of-edge state localized at the corner $x^{2} = x^3 = 0$ is
\begin{align}
 \psi = e^{-\alpha_2 x^2 - \alpha_3 x^3}
 \begin{pmatrix}
  \id_2 + i \sigma_{3} U_3
  \\ i \sigma_3 \left( \id_2 - i \sigma_3 U_3 \right)
 \end{pmatrix}
 \xi
\end{align}
with the compatibility condition
\begin{align}
 &
 \det
 \Big(
 \id_2 + i U_2^\dag \sigma_2 + i \sigma_3 U_3 - U_2^\dag (i\sigma_1) U_3
 \nonumber \\
 & \hspace{3em}
 + i \sigma_1 - i \sigma_2 U_3 - i U_2^\dag \sigma_3 - U_2^\dag U_3
 \Big) = 0
 \label{eq:UUcomp23}
\end{align}
since the boundary conditions \eqref{eq:bc_23} are rephrased as \eqref{eq:BC_re}.

A solution to the compatibility condition \eqref{eq:UUcomp23} is
\begin{align}
 U_2 = \sigma_2
 \, , \quad
 U_3 = i \id_2
 \, ,
 \label{eq:U2U3}
\end{align}
which leads to
\begin{align}
 (\tilde{p}^{(a)}_1, \tilde{p}^{(a)}_2, \tilde{p}^{(a)}_3, \tilde{p}^{(a)}_4)
 & =
 \begin{cases}
  (-p_3, p_4, p_1, p_5) & (a = 2) \\
  (p_1, p_2, -p_5, p_4) & (a = 3)
 \end{cases}
\end{align}
with $\theta_2 = \theta_3 = \pi/2$.
Thus the edge state spectrum is given by
\begin{subequations}
 \label{eq:edge_sp_23}
 \begin{align}
 \epsilon_2(p)
 & =
 \pm \sqrt{p_1^2 + p_3^2 + p_4^2}
 \, , \\
 \epsilon_3(p)
 & =
 \pm \sqrt{p_1^2 + p_2^2 + p_5^2}
 \, ,
\end{align}
\end{subequations}
and the corresponding edge-of-edge state spectrum is gapless and also chiral,
\begin{align}
 \epsilon = - p_1
 \, .
 \label{eq:eoe23_sp}
\end{align}

\subsection{3d class AIII topological insulator}

We consider the Hamiltonian for the class AIII topological insulator
\begin{align}
 \mathcal{H}(\vec{p})
 =
 \vec{p} \cdot \vec{\Gamma}
 + m \Gamma_4
 \, ,
 \label{Ham_AIII}
\end{align}
which is obtained from the 5d Weyl Hamiltonian \eqref{Hamil5} through the dimensional reduction $(p_4,p_5) \to (m,0)$.
We remark that the $\Gamma$-matrices \eqref{eq:Gamma_matrices} are expressed as
\begin{align}
 \Gamma^i = \tau_2 \otimes \sigma_i
 \, , \quad
 \Gamma^4 = \tau_1 \otimes \id_2
 \, , \quad
 \Gamma^5 = \tau_3 \otimes \id_2
\end{align}
where the Pauli matrices $\sigma$'s and $\tau$'s act on the spin $(\uparrow,\downarrow)$ and sublattice $(A,B)$ degrees of freedom.
Since the Hamiltonian anticommutes with $\Gamma^5$ as
\begin{align}
 \{\mathcal{H}(\vec{p}), \Gamma^5\} = 0 \, ,
 \label{eq:3d_chi_sym}
\end{align}
it has the chiral (sublattice) symmetry.

We can apply the same boundary analysis to the dimensionally reduced model.
Given a two-spinor denoted by $| \xi \rangle$, and choosing the boundary condition \eqref{eq:U2U3}, we obtain
\begin{align}
 \psi(x^2=0)
 \propto
 \begin{pmatrix}
  \id_2 \\ \sigma_2
 \end{pmatrix}
 |\xi\rangle
 \, , \quad
 \psi(x^3=0)
 \propto
 \begin{pmatrix}
  \id_2 - \sigma_3 \\ i (\id_2 + \sigma_3)
 \end{pmatrix}
 |\xi\rangle
 \, .
 \label{eq:3d_edge_state}
\end{align}
Since the operator $\id_2 \pm \sigma_3$ is a projector onto $\uparrow$ and $\downarrow$ spin state, we obtain the edge state $\psi(x^3=0)$ by applying $\downarrow$-spin projection to $A$ sites, and $\uparrow$-spin projection to $B$ sites at the $x^3=0$ plane.
On the other hand, another edge state $\psi(x^2 = 0)$ is obtained by applying the spin rotation generated by $\sigma_2$ only to $B$ site (nothing for $A$ site) at the $x^2=0$ plane.
The spectra of these boundary conditions are immediately obtained from \eqref{eq:edge_sp_23} with the reduction 
\begin{align}
 \epsilon_2(p)
 & = \pm
 \sqrt{p_1^2 + p_3^2 + m^2}
 \, , \quad
 \epsilon_3(p)
 = \pm
 \sqrt{p_1^2 + p_2^2}
 \, ,
 \label{eq:3d_edge_sp}
\end{align}
and the gapless edge-of-edge spectrum \eqref{eq:eoe23_sp}.
We now have the edge-of-edge state, but it seems difficult to detect its spectrum at this moment, because the spectrum $\epsilon_3(p)$ is also gapless in addition to the edge-of-edge state.
In order to distinguish the edge-of-edge state from the edge states, we need to consider the situation such that only the edge-of-edge state is gapless, while the other edge states are gapped.

Before studying such a situation, let us discuss the reason why either of the edge spectra \eqref{eq:3d_edge_sp} is gapless, while the other is gapped.
For the class AIII topological insulator, the gapless edge state is protected by the chiral symmetry \eqref{eq:3d_chi_sym}, which is indeed the sublattice symmetry.
However, if the boundary condition is not compatible with the symmetry which protects the topological property, the edge state cannot be gapless any longer.
This is essentially similar to the (class AII) topological insulator/ferromagnet junction~\cite{Qi:2008ew}.
The class AII topological insulator is protected by the time-reversal symmetry, but this symmetry can be weakly broken at the surface due to the junction with the ferromagnet.
The role of ferromagnet can be replaced by the chiral superconductor, which breaks the time-reversal symmetry~\cite{fu2008superconducting}.

From this point of view, the edge state at $x^2=0$ shown in \eqref{eq:3d_edge_state} breaks the sublattice symmetry because the $\sigma_2$-rotation acts only on the $B$-site, while the spin-projection applied to the edge state at $x^3=0$ could be consistent with the sublattice symmetry.
Thus, to gap out the spectrum $\epsilon_3(p)$, we need to explicitly break the chiral (sublattice) symmetry for the edge state at $x^3=0$.
For this purpose, we apply a rotated configuration
\begin{align}
 U_2 & =
 \sigma_2 \cos \phi + i \id_2 \sin \phi
 \, , \
 U_3 =
 i \id_2 \cos \phi - \sigma_3 \sin \phi
 \, ,
\end{align}
which satisfies the compatibility condition \eqref{eq:UUcomp23}.
Then we obtain the gapped edge spectra
\begin{align}
 \epsilon_2(p)
 & =
 \pm \sqrt{p_1^2 + p_3^2 + (m \cos \phi)^2}
 \, , \\
 \epsilon_3(p)
 & =
 \pm \sqrt{p_1^2 + p_2^2 + (m \sin \phi)^2}
 \, ,
\end{align}
with the edge-of-edge state \eqref{eq:eoe23_sp}.
Now only the edge-of-edge state is gapless, while the two edge states are gapped.
This could be a suitable situation for experimental detection of the edge-of-edge state.

\section{Topological charge of edge states}
\label{sec:top}

We show in subsection \ref{sec:mom} that the edge states of the 5D Weyl semimetal
have a topological charge identical to that of the bulks states of the 3D Weyl semimetals.
This makes sure the existence of the edge-of-edge state, since the edge-of-edge is seen
as a boundary of the edge surface which has a topological charge. Applying the
bulk-edge correspondence to the boundary (which is now interpreted as a bulk)
provides the existence of the edge-of-edge state.

In subsection \ref{sec:bout}, we point out that the Berry connection associated with the
edge state can be generalized to the space of the boundary conditions, not only the
space of the momentum. The Berry connection of the boundary condition space 
is shown to have a nontrivial Chern-Simons integral. The content of the
subsection is not directly related to the edge-of-edge states in the previous section.

\subsection{Topological charge in the momentum space}
\label{sec:mom}

In \cite{Hashimoto:2016dtm}, a certain edge state appearing in a class A topological insulator
in 1+4 dimensions was shown to possess a topological charge.
As argued earlier, we note here that the 5D Weyl semimetal Hamiltonian reduces
to a 4d class A topological insulator by a trivial dimensional reduction.
So, it is natural that our generic edge state explored in the previous section
has the same topological charge that was argued in \cite{Hashimoto:2016dtm}.

In fact, it is easy to see the topological charge of the edge state.
The topological charge is defined by a Berry connection of the wave function
of the edge state. Recall that the edge state wave function is subject to the
two equations \eqref{equi1} and \eqref{equi2}. In particular the second equation
\eqref{equi2} is recast to the form
\begin{align}
\sigma_i \tilde{p}_i\xi =
\left[\alpha_5\cos\theta_5+\epsilon\sin\theta_5\right]\xi 
 \, .
\end{align}
This is nothing but the Hamiltonian eigen equation for the 3D Weyl semimetal, \eqref{Hamil3D}.
Therefore, the Berry connection of the edge state has a topological charge.
It is identical to the chirality of the corresponding Weyl semimetal, in the
rotated momentum frame spanned by $\tilde{p}_{1,2,3}$.

The topological charge in the momentum space for the edge state immediately means
that there should appear an edge-of-edge state once a boundary of the edge is
introduced properly.

\subsection{Topological charge in the boundary condition space}
\label{sec:bout}

It was shown in our previous paper \cite{Hashimoto:2016kxm}
that the Berry connection of the edge state of the 3D Weyl semimetals
has a nontrivial topological structure. Since the parameters of
the edge states consist not only of the momenta but also of the parameter
of the boundary condition, the Berry connection associated with
the boundary condition space can be defined as well.
For the edge state \eqref{edgegeneral} with \eqref{alpha12}, its
Berry connection is calculated as
\begin{align}
A_{\theta_+} = 1 \, , \quad
A_{p_1}=A_{p_2}=0 \, .
\end{align}
Therefore the edge state has a nontrivial winding number along the space $\theta_+$ which 
parametrizes the boundary condition,
\begin{align}
\int_0^\pi d\theta_+ A_{\theta_+} = \pi \, .
\end{align}
This is a Wilson line, or in other words, a one-dimensional Chern-Simons action.

Let us see what will happen to our current case. The edge state wave function is now given as 
\begin{align}
\psi = \sqrt{\alpha_5} \; e^{-\alpha_5 x^5} 
\left(
\begin{array}{c}
\id_2
\\
e^{i\theta_5} U_5'
\end{array}
\right) \xi
\end{align}
with the two-spinor satisfying \eqref{equi1} and \eqref{equi2}, which means
\begin{align}
\left[\pm \sqrt{\tilde{p}_i^2}- \sigma_i \tilde{p}_i\right ] \xi = 0 \, ,
\end{align}
where $\tilde{p}_i$'s are defined by \eqref{tp} through $U_5'$. There are 
two edge states, specified by the $\pm$ sign. Explicitly, they are given by
\begin{align}
\xi_\pm = \frac{1}{\sqrt{2|\tilde{p}| (|\tilde{p}| \pm \tilde{p}_3)}}
\left(
\begin{array}{c}
\pm|\tilde{p}| + \tilde{p}_3
\\
\tilde{p}_1 + i\tilde{p}_2
\end{array}
\right) \, .
\label{pmxi}
\end{align}
The depth parameter $\alpha_5$
is given in \eqref{al5}, and the wave function is normalized as
\begin{align}
1 = \int_0^\infty dx^5 \; \psi^\dagger \psi \, ,
\end{align}
which is equivalent to $\xi^\dag \xi = 1$.
First, noting the relation to the 3D case, we easily find
\begin{align}
A_{\theta_5} \equiv
i \int_0^\infty dx^5 \; 
\psi^\dagger \frac{\partial}{\partial \theta_5}\psi
= -\frac12 
\end{align}
for each solutions corresponding to $\xi_\pm$.
So, again the edge state has a nontrivial topological structure in the $U(1)$
space of the boundary conditions spanned by $\theta_5$.
The Wilson line, or the one-dimensional Chern Simons term, is the same as the three-dimensional case,
\begin{align}
\int_0^{2\pi} d\theta_5 A_{\theta_5} = \pi \, .
\end{align}
For example, in the presence of a vortex surrounded by $S^1$, the Wilson line phase is
given by $2\pi$. So, compared with that, the present value $\pi$ is a half of 
a single winding. Therefore the wave function earns a phase $-1$ when
$\theta_5$ is rotated once around the boundary condition space.

Let us consider the Berry connection for
the $SU(2)$ part $U_5'$. The $SU(2)$ space 
is intertwined with the momentum space $\{p_i,p_4\}$ through
\eqref{tp}. However, in the new basis with $\{\tilde{p}_i, \tilde{p}_4\}$ they are decoupled with
each other. In this new basis $\{U_5', \tilde{p}_i, \tilde{p}_4\}$, the Berry connection is
calculated more easily. For the basis of the $SU(2)$ matrix $U_5'$, we choose
\begin{align}
U_5' = x_0 \id_2 + i \left(x_i \sigma_i \right)
\label{U5'x}
\end{align}
with $x_0^2 + x_i^2 = 1$. This is a three-sphere, so a canonical basis is the spherical coordinate
system,
\begin{align}
& x_0 = \cos\theta\, , 
\quad x_1 = \sin\theta \cos\phi \, , 
\nonumber
\\
& x_2 = \sin \theta \sin \phi \cos\chi \, , 
\quad x_3 = \sin \theta \sin \phi \sin\chi \, ,
\end{align}
with $0\leq \theta\leq \pi$, $0\leq\phi\leq\pi$ and $0\leq \chi\leq 2\pi$.
Using these coordinates, we can explicitly calculate the Berry connection
$(A_\theta, A_\phi, A_\chi)$.
The result is
\begin{align}
 A_a = \frac{i}{2} \,
 \xi_\pm^\dagger
 \left[
 (U_5')^\dagger \partial_a U_5'
 \right]
 \xi_\pm
 \, . 
\label{ourU5b}
\end{align}
Here the index $a$ runs for the spherical coordinates $(\theta, \phi, \chi)$.
Although this Berry connection $A_a$ still depends on the $SU(2)$-rotated momentum $\tilde{p}$ through $\xi_\pm$, the topological charge does not depend in the end, as shown below.
The typical topological charge on three-dimensional space is the Chern-Simons form,
\begin{align}
 \frac{1}{4\pi}
 \int d\theta d\phi d\chi \,
 \epsilon_{abc} \left(
 A_a \partial_b A_c\right)
 = \frac{\pi}{4}
 \label{cspi4}
\end{align}
So we find that the $SU(2)$ part of the boundary condition space
has a nontrivial topological structure.



The value $\pi/4$ of the Chern-Simons action in \eqref{cspi4} is $1/8$ of that for the single winding connection. The single winding connection of $S^3$ is
provided by an asymptotic connection of a BPST instanton, and is given
by 
\begin{align}
A_a = i U^\dagger \partial_a U \, ,
\label{single}
\end{align}
with $U = U_5'$ given by \eqref{U5'x}. 
It gives $S_{\rm CS} = 2\pi$. 
Notice that our Berry connection \eqref{ourU5b} is essentially $1/8$ of this
single-charge gauge connection \eqref{single}. This is the origin of the fact that our 
Chern-Simons action \eqref{cspi4} is $1/8$ of $2\pi$.


\section{Discussions}

In this paper we studied the 5-dimensional gapless Weyl fermion with
the Hamiltonian \eqref{Hamil5}, so the system looks non-realistic. However, a dimensional reduction leads to a three-dimensional chiral topological insulator of class AIII~\cite{Ryu:2010zza}.
For that case we have found an illuminating example where the
edge states are gapped while the edge-of-edge state is gapless.

The table of the classification of the topological phases has been studied
through dimensional reductions \cite{Ryu:2010zza}. We here introduced
another way to have a dimensional hierarchy: the intersection of the
boundary surfaces. In this paper we have just studied the two boundaries
meet at a right angle for simplicity, but in general, they need not to be.
The point is that when two boundaries with different boundary conditions
meet at codimension-2 surface, there could exist localized ``edge-of-edge'' states.
The topological charge in the boundary condition space studied in section \ref{sec:bout}
may characterize the existence condition, and we leave that question to our future problem.

It is important to mention the absence of the edge-of-edge state for the three-dimensional
Weyl semimetals. As seen in the generic boundary conditions of the 3D
Weyl semimetals \eqref{bc'}, they are given just by a single parameter $\theta_+$,
so a consistency condition \eqref{consUU} for the case of 5D cannot be
constructed for 3D. This follows from the fact that the Hamiltonian of the 3D case
is made of $2\times 2$ sigma matrices, while that of the 5D case is made of
$4\times 4$ Dirac matrices.  
So, in order to have the edge-of-edge state,
we needed to enhance the size of the Dirac operator by 2; 
the edge-of-edge state can exist when the Hamiltonian is given by $4\times 4$ 
matrix, basically the Dirac matrices in 4 or 5 dimensions.
From this argument, 
it is obvious that our argument could be generalized to $(2n+1)$-dimensional
Weyl semimetals ($n>2$), with a Hamiltonian of $2^n\times 2^n$ gamma matrices.
For that case, further possible localization, such as an edge-of-edge-of-edge,
is possible. In general, we can introduce $n$ edges with a completely localized state
at the intersection of all the edges. The construction is similar to the Atiyah-Bott-Shapiro
construction \cite{Atiyah:1964zz} of D-branes as a tachyon condensation of higher dimensional unstable D-branes
\cite{Witten:1998cd,Horava:1998jy,Kutasov:2000aq}.
We have just discussed class A and AIII examples, but its generalization to the system with time-reversal and particle-hole symmetries would be also possible~\cite{Hashimoto:2015dla}.


\begin{acknowledgments}
We would like to thank valuable discussions with H.~Fukaya, T.~Fukui, Y.~Hatsugai, N.~Kawakami, 
T.~Onogi and Y.~Tanaka.
The work of K.~H.~was supported in part by JSPS KAKENHI
Grant Number JP15H03658 and JP15K13483.
The work of T.~K.~was supported in part by Keio Gijuku Academic Development Funds, the MEXT-Supported Program for the Strategic Research Foundation at Private Universities ``Topological Science'' (No. S1511006), and JSPS Grant-in-Aid for Scientific Research on Innovative Areas ``Topological Materials Science'' (No. JP15H05855).
\end{acknowledgments}

\appendix
\section{Edge state at $x^{a}=0$ $(a\neq 5)$}
\label{sec:BC}

In this Appendix, we discuss the boundary conditions imposed to the 5d Weyl fermion \eqref{Hamil5} at the boundary $x^a=0$ for $a = 1, 2, 3, 4$ in details.
The boundary condition which we consider is
\begin{align}
 \psi^\dag \Gamma^a \psi\Big|_{x^{a} = 0} = 0
 \quad (a = 1,2,3,4)
 \, .
 \label{eq:bc_1234}
\end{align}
We use the notation of the $\Gamma$-matrices shown in \eqref{eq:Gamma_matrices}.
Applying the same argument discussed in Sec.~\ref{sec:5D}, we obtain the corresponding localized edge state
\begin{align}
 (a = 1, 2, 3): \ \psi
 & =
 e^{- \alpha_a x^a}
 \begin{pmatrix}
  \id_2 + i \sigma_{a} U_{a}
  \\ i \sigma_a \left( \id_2 - i \sigma_a U_a \right)
 \end{pmatrix} 
 \xi_{a}
 \, ,
 \nonumber \\
 (a = 4): \
 \psi
 & =
 e^{-\alpha_4 x^4}
 \begin{pmatrix}
  \id_2 - U_4 \\ \id_2 + U_4
 \end{pmatrix}
 \xi_4
 \, ,
\end{align}
with $U_a \in U(2)$ and a two-spinor $\xi_a$.
We remark that the boundary conditions \eqref{eq:bc_1234} are rephrased as
\begin{align}
 \begin{pmatrix}
  \id_2 + i U^\dag_a \sigma_a &
  \left( \id_2 - i U^\dag_a \sigma_a \right) (-i\sigma_a)
 \end{pmatrix}
 \psi\Big|_{x^a=0}
 = 0
 \label{eq:BC_re}
\end{align}
for $a = 1,2,3$, and
\begin{align}
 \begin{pmatrix}
  \id_2 + U^\dag_4 & \id_2 - U^\dag_4
 \end{pmatrix}
 \psi\Big|_{x^4=0}
 = 0
\end{align}
for $a = 4$.

Let us then solve the spectrum of the edge state localized at $x^a = 0$.
The eigen equation $\mathcal{H} \psi = \epsilon \psi$ for the Hamiltonian \eqref{Hamil5} with the boundary condition \eqref{eq:bc_1234} for $a = 1$ leads to
\begin{subequations} 
 \begin{align}
  \Big(
  \left( i \alpha_1 - \epsilon \right)
  + (i p_5 \sigma_1 - i p_2 \sigma_2 - i p_3 \sigma_3 + p_4) U_1
  \Big) \xi_1
  & = 0
  \, , \\
  \Big(
  - \left( i \alpha_1 + \epsilon \right) U_1
  + (- i p_5 \sigma_1 + i p_2 \sigma_2 + i p_3 \sigma_3 + p_4) 
  \Big) \xi_1
  & = 0
  \, .
 \end{align}
\end{subequations}
We can similarly discuss the edge state localized at $x^2 = 0$, $x^3 = 0$, and also for $x^4 = 0$, which lead to the condition identical to that studied in Sec.~\ref{sec:5D} if we replace
\begin{align}
 (\alpha_a, p_5)
 \ \longrightarrow \
 (\alpha_5, -p_a)
 \qquad
 (a = 1, 2, 3, 4)
 \, .
\end{align}
Thus we obtain
\begin{align}
 \epsilon_a(p)
 & =
 p_4^{(a)} \cos \theta_a
 \pm \sqrt{ |p_i^{(a)}|^2 } \sin \theta_a
 \, , \\
 \alpha_a(p)
 & =
 - p_4^{(a)} \sin \theta_a
 \pm \sqrt{ |p_i^{(a)}|^2 } \cos \theta_a 
\end{align}
where we decompose $U_a = e^{i\theta_a} U'_a$ with $U'_a \in SU(2)$, and define
\begin{align}
 (p^{(a)}_1, p^{(a)}_2, p^{(a)}_3, p^{(a)}_4)
 & =
 \begin{cases}
  (-p_5, p_2, p_3, p_4) & (a = 1) \\
  (p_1, -p_5, p_3, p_4) & (a = 2) \\  
  (p_1, p_2, -p_5, p_4) & (a = 3) \\
  (p_1, p_2, p_3, -p_5) & (a = 4) \\  
 \end{cases}
\end{align}
with the $SU(2)$-rotated momentum for $a = 1,2,3,4$,
\begin{align}
 ( - i p^{(a)}_i \sigma_i + p^{(a)}_4) U'_a
 & =
 - i \tilde{p}^{(a)}_i \sigma_i + \tilde{p}^{(a)}_4
 \, .
\end{align}
The normalization condition of the edge state is $\alpha_a > 0$.

\section{Derivation of generic edge-of-edge dispersion relation}
\label{sec:der_eoe}

We show the derivation of the dispersion relation of generic edge-of-edge state localized at $x^4=x^5=0$.
Parametrizing the boundary condition matrices \eqref{eq:U4U5_parametrization}, the compatibility condition \eqref{consUU} becomes
\begin{align}
 &\det \big[\id_2+B_0-A_0+A_{\mu}B^{\mu}+ \nonumber
 \\
 &(B_i-A_i+A_iB_0+A_0B_i+i\epsilon_{ijk}A_jB_k)\sigma^i\big]=0\, ,
\end{align}
which can be further written as:
\begin{align}
 \id_2+2(B_0-A_0)+A_0^2-A_i^2+B_0^2-B_i^2+2A_0(B_0^2-B_i^2) \nonumber
 \\
 -2B_0(A_0^2-A_i^2)+(A_0^2-A_i^2)(B_0^2-B_i^2)+4A_iB_i=0.
\end{align}
Using the fact that $A_0^2-A_i^2=e^{2i\theta_5}$ and $B_0^2-B_i^2=e^{2i\theta_4}$, 
this is shown to be equivalent to  
\begin{align}	
	a_ib_i=-\cos \theta_4 \cos \theta_5-ia_0\sin \theta_4+ib_0\sin \theta_5\, ,
\end{align}
and we arrive at the following two equations:
\begin{empheq}[left=\empheqlbrace]{align}
		&a_ib_i=-\cos \theta_4 \cos \theta_5\, ,
		\\  
		&a_0\sin \theta_4=b_0\sin \theta_5\, .
\end{empheq}
This is the generic constraint for the two boundary conditions, for the existence
of the edge-of-edge states.

Next let us solve the energy eigen equations, \eqref{a1--} and \eqref{a2--}. Denoting
\begin{align}
	p_5:=i\alpha_5 \label{pa5}
\end{align}
and also
\begin{align}
	(i\sigma_jp_j+p_5)(b_0+ib_i\sigma_i)=i\sigma_i\tilde{\tilde{p}}_i+\tilde{\tilde{p}}_5\, ,
\end{align}
we have
\begin{empheq}[left=\empheqlbrace]{align}
	&\tilde{\tilde{p}}_5=b_0p_5-b_ip_i
	\\
	&\tilde{\tilde{p}}_i=b_0p_i+b_ip_5+\epsilon_{ijk}b_jp_k\, .
\end{empheq}
Then equations \eqref{a1--} and \eqref{a2--} become
\begin{empheq}[left=\empheqlbrace]{align}
	&\epsilon \cos\theta_4-\alpha_4\sin \theta_4+\tilde{\tilde{p}}_5=0\, ,
	\\
	&(\epsilon \sin\theta_4+\alpha_4\cos\theta_4)^2-\tilde{\tilde{p}}_i^2=0\, .
\end{empheq}
These two equations are related by
$	\epsilon^2=\tilde{\tilde{p}}_i^2+\tilde{\tilde{p}}_5^2-\alpha_4^2$, so 
instead, we shall use the following equivalent set of equations, 
\begin{empheq}[left=\empheqlbrace]{align}
	&\epsilon \cos\theta_4-\alpha_4\sin \theta_4=b_ip_i-b_0p_5\, ,
	\\
	&\epsilon^2=p_i^2-\alpha_4^2-\alpha_5^2 \,  \label{epaa}
\end{empheq}
for convenience.
Since \eqref{pa5} means that $p_5$ is pure imaginary, 
above two equations are actually three real equations including
 $	b_0p_5=0$, 
 which means 
  \begin{align}
 	b_0=0\, ,
 \end{align}
 and 
 \begin{align}\label{disp4}
 	\epsilon \cos\theta_4-\alpha_4\sin \theta_4=b_ip_i . 
 \end{align}

Similarly, consider the boundary condition on the $x^5$ direction. Substitute equation \eqref{bou55} into 
the energy eigen equation and repeat the procedures starting from equations \eqref{a1--} and \eqref{a2--}. 
Then we obtain
 \begin{empheq}[left=\empheqlbrace]{align}
	&\epsilon \cos\theta_5-\alpha_5\sin \theta_5=a_ip_i\, , \label{disp5}
	\\ &a_0=0 \, .
\end{empheq}
 Combining equations \eqref{disp4} \eqref{disp5} and  \eqref{epaa} to eliminate $\alpha_4$ and $\alpha_5$,
 we obtain
  \begin{align}
   A\epsilon^2-2B\epsilon+C=0\, ,
  \end{align}
which is \eqref{eq:eoe_disp} with the coefficients defined in \eqref{eq:ABC}.

\bibliography{3DTI-revtex}

\begin{thebibliography}{35}%
\makeatletter
\providecommand \@ifxundefined [1]{%
 \@ifx{#1\undefined}
}%
\providecommand \@ifnum [1]{%
 \ifnum #1\expandafter \@firstoftwo
 \else \expandafter \@secondoftwo
 \fi
}%
\providecommand \@ifx [1]{%
 \ifx #1\expandafter \@firstoftwo
 \else \expandafter \@secondoftwo
 \fi
}%
\providecommand \natexlab [1]{#1}%
\providecommand \enquote  [1]{``#1''}%
\providecommand \bibnamefont  [1]{#1}%
\providecommand \bibfnamefont [1]{#1}%
\providecommand \citenamefont [1]{#1}%
\providecommand \href@noop [0]{\@secondoftwo}%
\providecommand \href [0]{\begingroup \@sanitize@url \@href}%
\providecommand \@href[1]{\@@startlink{#1}\@@href}%
\providecommand \@@href[1]{\endgroup#1\@@endlink}%
\providecommand \@sanitize@url [0]{\catcode `\\12\catcode `\$12\catcode
  `\&12\catcode `\#12\catcode `\^12\catcode `\_12\catcode `\%12\relax}%
\providecommand \@@startlink[1]{}%
\providecommand \@@endlink[0]{}%
\providecommand \url  [0]{\begingroup\@sanitize@url \@url }%
\providecommand \@url [1]{\endgroup\@href {#1}{\urlprefix }}%
\providecommand \urlprefix  [0]{URL }%
\providecommand \Eprint [0]{\href }%
\providecommand \doibase [0]{http://dx.doi.org/}%
\providecommand \selectlanguage [0]{\@gobble}%
\providecommand \bibinfo  [0]{\@secondoftwo}%
\providecommand \bibfield  [0]{\@secondoftwo}%
\providecommand \translation [1]{[#1]}%
\providecommand \BibitemOpen [0]{}%
\providecommand \bibitemStop [0]{}%
\providecommand \bibitemNoStop [0]{.\EOS\space}%
\providecommand \EOS [0]{\spacefactor3000\relax}%
\providecommand \BibitemShut  [1]{\csname bibitem#1\endcsname}%
\let\auto@bib@innerbib\@empty
\bibitem [{\citenamefont {Jackiw}\ and\ \citenamefont
  {Rebbi}(1976)}]{Jackiw:1975fn}%
  \BibitemOpen
  \bibfield  {author} {\bibinfo {author} {\bibfnamefont {R.}~\bibnamefont
  {Jackiw}}\ and\ \bibinfo {author} {\bibfnamefont {C.}~\bibnamefont {Rebbi}},\
  }\href {\doibase 10.1103/PhysRevD.13.3398} {\bibfield  {journal} {\bibinfo
  {journal} {Phys. Rev.}\ }\textbf {\bibinfo {volume} {D13}},\ \bibinfo {pages}
  {3398} (\bibinfo {year} {1976})}\BibitemShut {NoStop}%
\bibitem [{\citenamefont {Hatsugai}(1993)}]{hatsugai1993chern}%
  \BibitemOpen
  \bibfield  {author} {\bibinfo {author} {\bibfnamefont {Y.}~\bibnamefont
  {Hatsugai}},\ }\href {\doibase 10.1103/PhysRevLett.71.3697} {\bibfield
  {journal} {\bibinfo  {journal} {Phys. Rev. Lett.}\ }\textbf {\bibinfo
  {volume} {71}},\ \bibinfo {pages} {3697} (\bibinfo {year}
  {1993})}\BibitemShut {NoStop}%
\bibitem [{\citenamefont {Wen}(2004)}]{Wen:2004ym}%
  \BibitemOpen
  \bibfield  {author} {\bibinfo {author} {\bibfnamefont {X.-G.}\ \bibnamefont
  {Wen}},\ }\href {\doibase 10.1093/acprof:oso/9780199227259.001.0001} {\emph
  {\bibinfo {title} {{Quantum Field Theory of Many-Body Systems: From the
  Origin of Sound to an Origin of Light and Electrons}}}}\ (\bibinfo
  {publisher} {Oxford Univ. Press},\ \bibinfo {year} {2004})\BibitemShut
  {NoStop}%
\bibitem [{\citenamefont {Hasan}\ and\ \citenamefont
  {Kane}(2010)}]{hasan2010colloquium}%
  \BibitemOpen
  \bibfield  {author} {\bibinfo {author} {\bibfnamefont {M.~Z.}\ \bibnamefont
  {Hasan}}\ and\ \bibinfo {author} {\bibfnamefont {C.~L.}\ \bibnamefont
  {Kane}},\ }\href {\doibase 10.1103/RevModPhys.82.3045} {\bibfield  {journal}
  {\bibinfo  {journal} {Rev. Mod. Phys.}\ }\textbf {\bibinfo {volume} {82}},\
  \bibinfo {pages} {3045} (\bibinfo {year} {2010})},\ \Eprint
  {http://arxiv.org/abs/1002.3895} {arXiv:1002.3895 [cond-mat.mes-hall]}
  \BibitemShut {NoStop}%
\bibitem [{\citenamefont {Qi}\ and\ \citenamefont
  {Zhang}(2011)}]{qi2011topological}%
  \BibitemOpen
  \bibfield  {author} {\bibinfo {author} {\bibfnamefont {X.-L.}\ \bibnamefont
  {Qi}}\ and\ \bibinfo {author} {\bibfnamefont {S.-C.}\ \bibnamefont {Zhang}},\
  }\href {\doibase 10.1103/RevModPhys.83.1057} {\bibfield  {journal} {\bibinfo
  {journal} {Rev. Mod. Phys.}\ }\textbf {\bibinfo {volume} {83}},\ \bibinfo
  {pages} {1057} (\bibinfo {year} {2011})},\ \Eprint
  {http://arxiv.org/abs/1008.2026} {arXiv:1008.2026 [cond-mat.mes-hall]}
  \BibitemShut {NoStop}%
\bibitem [{\citenamefont {Witten}(1998)}]{Witten:1998cd}%
  \BibitemOpen
  \bibfield  {author} {\bibinfo {author} {\bibfnamefont {E.}~\bibnamefont
  {Witten}},\ }\href {\doibase 10.1088/1126-6708/1998/12/019} {\bibfield
  {journal} {\bibinfo  {journal} {JHEP}\ }\textbf {\bibinfo {volume} {12}},\
  \bibinfo {pages} {019} (\bibinfo {year} {1998})},\ \Eprint
  {http://arxiv.org/abs/hep-th/9810188} {arXiv:hep-th/9810188 [hep-th]}
  \BibitemShut {NoStop}%
\bibitem [{\citenamefont {Schnyder}\ \emph {et~al.}(2008)\citenamefont
  {Schnyder}, \citenamefont {Ryu}, \citenamefont {Furusaki},\ and\
  \citenamefont {Ludwig}}]{Schnyder:2008tya}%
  \BibitemOpen
  \bibfield  {author} {\bibinfo {author} {\bibfnamefont {A.~P.}\ \bibnamefont
  {Schnyder}}, \bibinfo {author} {\bibfnamefont {S.}~\bibnamefont {Ryu}},
  \bibinfo {author} {\bibfnamefont {A.}~\bibnamefont {Furusaki}}, \ and\
  \bibinfo {author} {\bibfnamefont {A.~W.~W.}\ \bibnamefont {Ludwig}},\ }\href
  {\doibase 10.1103/PhysRevB.78.195125} {\bibfield  {journal} {\bibinfo
  {journal} {Phys. Rev.}\ }\textbf {\bibinfo {volume} {B78}},\ \bibinfo {pages}
  {195125} (\bibinfo {year} {2008})},\ \Eprint {http://arxiv.org/abs/0803.2786}
  {arXiv:0803.2786 [cond-mat.mes-hall]} \BibitemShut {NoStop}%
\bibitem [{\citenamefont {Kitaev}(2009)}]{Kitaev:2009mg}%
  \BibitemOpen
  \bibfield  {author} {\bibinfo {author} {\bibfnamefont {A.}~\bibnamefont
  {Kitaev}},\ }\bibfield  {booktitle} {\emph {\bibinfo {booktitle} {{Advances
  in theoretical physics. Proceedings, Landau Memorial Conference,
  Chernogolokova, Russia, June 22-26, 2008}}},\ }\href {\doibase
  10.1063/1.3149495} {\bibfield  {journal} {\bibinfo  {journal} {AIP Conf.
  Proc.}\ }\textbf {\bibinfo {volume} {1134}},\ \bibinfo {pages} {22} (\bibinfo
  {year} {2009})},\ \Eprint {http://arxiv.org/abs/0901.2686} {arXiv:0901.2686
  [cond-mat.mes-hall]} \BibitemShut {NoStop}%
\bibitem [{\citenamefont {{Xu}}\ \emph {et~al.}(2015)\citenamefont {{Xu}},
  \citenamefont {{Belopolski}}, \citenamefont {{Alidoust}}, \citenamefont
  {{Neupane}}, \citenamefont {{Bian}}, \citenamefont {{Zhang}}, \citenamefont
  {{Sankar}}, \citenamefont {{Chang}}, \citenamefont {{Yuan}}, \citenamefont
  {{Lee}}, \citenamefont {{Huang}}, \citenamefont {{Zheng}}, \citenamefont
  {{Ma}}, \citenamefont {{Sanchez}}, \citenamefont {{Wang}}, \citenamefont
  {{Bansil}}, \citenamefont {{Chou}}, \citenamefont {{Shibayev}}, \citenamefont
  {{Lin}}, \citenamefont {{Jia}},\ and\ \citenamefont
  {{Hasan}}}]{Xu:2015Science}%
  \BibitemOpen
  \bibfield  {author} {\bibinfo {author} {\bibfnamefont {S.-Y.}\ \bibnamefont
  {{Xu}}}, \bibinfo {author} {\bibfnamefont {I.}~\bibnamefont {{Belopolski}}},
  \bibinfo {author} {\bibfnamefont {N.}~\bibnamefont {{Alidoust}}}, \bibinfo
  {author} {\bibfnamefont {M.}~\bibnamefont {{Neupane}}}, \bibinfo {author}
  {\bibfnamefont {G.}~\bibnamefont {{Bian}}}, \bibinfo {author} {\bibfnamefont
  {C.}~\bibnamefont {{Zhang}}}, \bibinfo {author} {\bibfnamefont
  {R.}~\bibnamefont {{Sankar}}}, \bibinfo {author} {\bibfnamefont
  {G.}~\bibnamefont {{Chang}}}, \bibinfo {author} {\bibfnamefont
  {Z.}~\bibnamefont {{Yuan}}}, \bibinfo {author} {\bibfnamefont {C.-C.}\
  \bibnamefont {{Lee}}}, \bibinfo {author} {\bibfnamefont {S.-M.}\ \bibnamefont
  {{Huang}}}, \bibinfo {author} {\bibfnamefont {H.}~\bibnamefont {{Zheng}}},
  \bibinfo {author} {\bibfnamefont {J.}~\bibnamefont {{Ma}}}, \bibinfo {author}
  {\bibfnamefont {D.~S.}\ \bibnamefont {{Sanchez}}}, \bibinfo {author}
  {\bibfnamefont {B.}~\bibnamefont {{Wang}}}, \bibinfo {author} {\bibfnamefont
  {A.}~\bibnamefont {{Bansil}}}, \bibinfo {author} {\bibfnamefont
  {F.}~\bibnamefont {{Chou}}}, \bibinfo {author} {\bibfnamefont {P.~P.}\
  \bibnamefont {{Shibayev}}}, \bibinfo {author} {\bibfnamefont
  {H.}~\bibnamefont {{Lin}}}, \bibinfo {author} {\bibfnamefont
  {S.}~\bibnamefont {{Jia}}}, \ and\ \bibinfo {author} {\bibfnamefont {M.~Z.}\
  \bibnamefont {{Hasan}}},\ }\href {\doibase 10.1126/science.aaa9297}
  {\bibfield  {journal} {\bibinfo  {journal} {Science}\ }\textbf {\bibinfo
  {volume} {349}},\ \bibinfo {pages} {613} (\bibinfo {year} {2015})},\ \Eprint
  {http://arxiv.org/abs/1502.03807} {arXiv:1502.03807 [cond-mat.mes-hall]}
  \BibitemShut {NoStop}%
\bibitem [{\citenamefont {Huang}\ \emph {et~al.}(2015)\citenamefont {Huang},
  \citenamefont {Xu}, \citenamefont {Belopolski}, \citenamefont {Lee},
  \citenamefont {Chang}, \citenamefont {Wang}, \citenamefont {Alidoust},
  \citenamefont {Bian}, \citenamefont {Neupane}, \citenamefont {Zhang},
  \citenamefont {Jia}, \citenamefont {Bansil}, \citenamefont {Lin},\ and\
  \citenamefont {Hasan}}]{Huang:2015NC}%
  \BibitemOpen
  \bibfield  {author} {\bibinfo {author} {\bibfnamefont {S.-M.}\ \bibnamefont
  {Huang}}, \bibinfo {author} {\bibfnamefont {S.-Y.}\ \bibnamefont {Xu}},
  \bibinfo {author} {\bibfnamefont {I.}~\bibnamefont {Belopolski}}, \bibinfo
  {author} {\bibfnamefont {C.-C.}\ \bibnamefont {Lee}}, \bibinfo {author}
  {\bibfnamefont {G.}~\bibnamefont {Chang}}, \bibinfo {author} {\bibfnamefont
  {B.}~\bibnamefont {Wang}}, \bibinfo {author} {\bibfnamefont {N.}~\bibnamefont
  {Alidoust}}, \bibinfo {author} {\bibfnamefont {G.}~\bibnamefont {Bian}},
  \bibinfo {author} {\bibfnamefont {M.}~\bibnamefont {Neupane}}, \bibinfo
  {author} {\bibfnamefont {C.}~\bibnamefont {Zhang}}, \bibinfo {author}
  {\bibfnamefont {S.}~\bibnamefont {Jia}}, \bibinfo {author} {\bibfnamefont
  {A.}~\bibnamefont {Bansil}}, \bibinfo {author} {\bibfnamefont
  {H.}~\bibnamefont {Lin}}, \ and\ \bibinfo {author} {\bibfnamefont {M.~Z.}\
  \bibnamefont {Hasan}},\ }\href {\doibase 10.1038/ncomms8373} {\bibfield
  {journal} {\bibinfo  {journal} {Nature Comm.}\ }\textbf {\bibinfo {volume}
  {6}},\ \bibinfo {pages} {7373} (\bibinfo {year} {2015})}\BibitemShut
  {NoStop}%
\bibitem [{\citenamefont {Weng}\ \emph {et~al.}(2015)\citenamefont {Weng},
  \citenamefont {Fang}, \citenamefont {Fang}, \citenamefont {Bernevig},\ and\
  \citenamefont {Dai}}]{Weng2015:PRX}%
  \BibitemOpen
  \bibfield  {author} {\bibinfo {author} {\bibfnamefont {H.}~\bibnamefont
  {Weng}}, \bibinfo {author} {\bibfnamefont {C.}~\bibnamefont {Fang}}, \bibinfo
  {author} {\bibfnamefont {Z.}~\bibnamefont {Fang}}, \bibinfo {author}
  {\bibfnamefont {B.~A.}\ \bibnamefont {Bernevig}}, \ and\ \bibinfo {author}
  {\bibfnamefont {X.}~\bibnamefont {Dai}},\ }\href {\doibase
  10.1103/PhysRevX.5.011029} {\bibfield  {journal} {\bibinfo  {journal} {Phys.
  Rev.}\ }\textbf {\bibinfo {volume} {X5}},\ \bibinfo {pages} {011029}
  (\bibinfo {year} {2015})},\ \Eprint {http://arxiv.org/abs/1501.00060}
  {arXiv:1501.00060 [cond-mat.mtrl-sci]} \BibitemShut {NoStop}%
\bibitem [{\citenamefont {Murakami}\ \emph {et~al.}(2007)\citenamefont
  {Murakami}, \citenamefont {Iso}, \citenamefont {Avishai}, \citenamefont
  {Onoda},\ and\ \citenamefont {Nagaosa}}]{Murakami:2007bx}%
  \BibitemOpen
  \bibfield  {author} {\bibinfo {author} {\bibfnamefont {S.}~\bibnamefont
  {Murakami}}, \bibinfo {author} {\bibfnamefont {S.}~\bibnamefont {Iso}},
  \bibinfo {author} {\bibfnamefont {Y.}~\bibnamefont {Avishai}}, \bibinfo
  {author} {\bibfnamefont {M.}~\bibnamefont {Onoda}}, \ and\ \bibinfo {author}
  {\bibfnamefont {N.}~\bibnamefont {Nagaosa}},\ }\href {\doibase
  10.1103/PhysRevB.76.205304} {\bibfield  {journal} {\bibinfo  {journal} {Phys.
  Rev.}\ }\textbf {\bibinfo {volume} {B76}},\ \bibinfo {pages} {205304}
  (\bibinfo {year} {2007})},\ \Eprint {http://arxiv.org/abs/0705.3696}
  {arXiv:0705.3696 [cond-mat.mes-hall]} \BibitemShut {NoStop}%
\bibitem [{\citenamefont {Murakami}(2007)}]{Murakami:2007NJP}%
  \BibitemOpen
  \bibfield  {author} {\bibinfo {author} {\bibfnamefont {S.}~\bibnamefont
  {Murakami}},\ }\href {\doibase 10.1088/1367-2630/9/9/356} {\bibfield
  {journal} {\bibinfo  {journal} {New J. Phys.}\ }\textbf {\bibinfo {volume}
  {9}},\ \bibinfo {pages} {356} (\bibinfo {year} {2007})},\ \Eprint
  {http://arxiv.org/abs/0710.0930} {arXiv:0710.0930 [cond-mat.mes-hall]}
  \BibitemShut {NoStop}%
\bibitem [{\citenamefont {Wan}\ \emph {et~al.}(2011)\citenamefont {Wan},
  \citenamefont {Turner}, \citenamefont {Vishwanath},\ and\ \citenamefont
  {Savrasov}}]{Wan:2011PRB}%
  \BibitemOpen
  \bibfield  {author} {\bibinfo {author} {\bibfnamefont {X.}~\bibnamefont
  {Wan}}, \bibinfo {author} {\bibfnamefont {A.~M.}\ \bibnamefont {Turner}},
  \bibinfo {author} {\bibfnamefont {A.}~\bibnamefont {Vishwanath}}, \ and\
  \bibinfo {author} {\bibfnamefont {S.~Y.}\ \bibnamefont {Savrasov}},\ }\href
  {\doibase 10.1103/PhysRevB.83.205101} {\bibfield  {journal} {\bibinfo
  {journal} {Phys. Rev.}\ }\textbf {\bibinfo {volume} {B83}},\ \bibinfo {pages}
  {205101} (\bibinfo {year} {2011})},\ \Eprint {http://arxiv.org/abs/1007.0016}
  {arXiv:1007.0016 [cond-mat.str-el]} \BibitemShut {NoStop}%
\bibitem [{\citenamefont {Yang}\ \emph {et~al.}(2011)\citenamefont {Yang},
  \citenamefont {Lu},\ and\ \citenamefont {Ran}}]{Yang:2011PRB}%
  \BibitemOpen
  \bibfield  {author} {\bibinfo {author} {\bibfnamefont {K.-Y.}\ \bibnamefont
  {Yang}}, \bibinfo {author} {\bibfnamefont {Y.-M.}\ \bibnamefont {Lu}}, \ and\
  \bibinfo {author} {\bibfnamefont {Y.}~\bibnamefont {Ran}},\ }\href {\doibase
  10.1103/PhysRevB.84.075129} {\bibfield  {journal} {\bibinfo  {journal} {Phys.
  Rev.}\ }\textbf {\bibinfo {volume} {B84}},\ \bibinfo {pages} {075129}
  (\bibinfo {year} {2011})},\ \Eprint {http://arxiv.org/abs/1105.2353}
  {arXiv:1105.2353 [cond-mat.str-el]} \BibitemShut {NoStop}%
\bibitem [{\citenamefont {Burkov}\ and\ \citenamefont
  {Balents}(2011)}]{Burkov2011:PRL}%
  \BibitemOpen
  \bibfield  {author} {\bibinfo {author} {\bibfnamefont {A.~A.}\ \bibnamefont
  {Burkov}}\ and\ \bibinfo {author} {\bibfnamefont {L.}~\bibnamefont
  {Balents}},\ }\href {\doibase 10.1103/PhysRevLett.107.127205} {\bibfield
  {journal} {\bibinfo  {journal} {Phys. Rev. Lett.}\ }\textbf {\bibinfo
  {volume} {107}},\ \bibinfo {pages} {127205} (\bibinfo {year} {2011})},\
  \Eprint {http://arxiv.org/abs/1105.5138} {arXiv:1105.5138
  [cond-mat.mes-hall]} \BibitemShut {NoStop}%
\bibitem [{\citenamefont {Xu}\ \emph {et~al.}(2011)\citenamefont {Xu},
  \citenamefont {Weng}, \citenamefont {Wang}, \citenamefont {Dai},\ and\
  \citenamefont {Fang}}]{Xu:2011dn}%
  \BibitemOpen
  \bibfield  {author} {\bibinfo {author} {\bibfnamefont {G.}~\bibnamefont
  {Xu}}, \bibinfo {author} {\bibfnamefont {H.}~\bibnamefont {Weng}}, \bibinfo
  {author} {\bibfnamefont {Z.}~\bibnamefont {Wang}}, \bibinfo {author}
  {\bibfnamefont {X.}~\bibnamefont {Dai}}, \ and\ \bibinfo {author}
  {\bibfnamefont {Z.}~\bibnamefont {Fang}},\ }\href {\doibase
  10.1103/PhysRevLett.107.186806} {\bibfield  {journal} {\bibinfo  {journal}
  {Phys. Rev. Lett.}\ }\textbf {\bibinfo {volume} {107}},\ \bibinfo {pages}
  {186806} (\bibinfo {year} {2011})},\ \Eprint {http://arxiv.org/abs/1106.3125}
  {arXiv:1106.3125 [cond-mat.mes-hall]} \BibitemShut {NoStop}%
\bibitem [{\citenamefont {Burkov}\ \emph {et~al.}(2011)\citenamefont {Burkov},
  \citenamefont {Hook},\ and\ \citenamefont {Balents}}]{Burkov2011:PRB}%
  \BibitemOpen
  \bibfield  {author} {\bibinfo {author} {\bibfnamefont {A.~A.}\ \bibnamefont
  {Burkov}}, \bibinfo {author} {\bibfnamefont {M.~D.}\ \bibnamefont {Hook}}, \
  and\ \bibinfo {author} {\bibfnamefont {L.}~\bibnamefont {Balents}},\ }\href
  {\doibase 10.1103/PhysRevB.84.235126} {\bibfield  {journal} {\bibinfo
  {journal} {Phys. Rev.}\ }\textbf {\bibinfo {volume} {B84}},\ \bibinfo {pages}
  {235126} (\bibinfo {year} {2011})},\ \Eprint {http://arxiv.org/abs/1110.1089}
  {arXiv:1110.1089 [cond-mat.mes-hall]} \BibitemShut {NoStop}%
\bibitem [{\citenamefont {Hashimoto}\ \emph {et~al.}(2016)\citenamefont
  {Hashimoto}, \citenamefont {Kimura},\ and\ \citenamefont
  {Wu}}]{Hashimoto:2016kxm}%
  \BibitemOpen
  \bibfield  {author} {\bibinfo {author} {\bibfnamefont {K.}~\bibnamefont
  {Hashimoto}}, \bibinfo {author} {\bibfnamefont {T.}~\bibnamefont {Kimura}}, \
  and\ \bibinfo {author} {\bibfnamefont {X.}~\bibnamefont {Wu}},\ }\href@noop
  {} {\  (\bibinfo {year} {2016})},\ \Eprint {http://arxiv.org/abs/1609.00884}
  {arXiv:1609.00884 [cond-mat.mes-hall]} \BibitemShut {NoStop}%
\bibitem [{\citenamefont {Isaev}\ \emph {et~al.}(2011)\citenamefont {Isaev},
  \citenamefont {Moon},\ and\ \citenamefont {Ortiz}}]{isaev2011bulk}%
  \BibitemOpen
  \bibfield  {author} {\bibinfo {author} {\bibfnamefont {L.}~\bibnamefont
  {Isaev}}, \bibinfo {author} {\bibfnamefont {Y.~H.}\ \bibnamefont {Moon}}, \
  and\ \bibinfo {author} {\bibfnamefont {G.}~\bibnamefont {Ortiz}},\ }\href
  {\doibase 10.1103/PhysRevB.84.075444} {\bibfield  {journal} {\bibinfo
  {journal} {Phys. Rev.}\ }\textbf {\bibinfo {volume} {B84}},\ \bibinfo {pages}
  {075444} (\bibinfo {year} {2011})},\ \Eprint {http://arxiv.org/abs/1103.0025}
  {arXiv:1103.0025 [cond-mat.mes-hall]} \BibitemShut {NoStop}%
\bibitem [{\citenamefont {Enaldiev}\ \emph {et~al.}(2015)\citenamefont
  {Enaldiev}, \citenamefont {Zagorodnev},\ and\ \citenamefont
  {Volkov}}]{Enaldiev:2015JETP}%
  \BibitemOpen
  \bibfield  {author} {\bibinfo {author} {\bibfnamefont {V.~V.}\ \bibnamefont
  {Enaldiev}}, \bibinfo {author} {\bibfnamefont {I.~V.}\ \bibnamefont
  {Zagorodnev}}, \ and\ \bibinfo {author} {\bibfnamefont {V.~A.}\ \bibnamefont
  {Volkov}},\ }\href {\doibase 10.1134/S0021364015020071} {\bibfield  {journal}
  {\bibinfo  {journal} {JETP Lett.}\ }\textbf {\bibinfo {volume} {101}},\
  \bibinfo {pages} {89} (\bibinfo {year} {2015})},\ \Eprint
  {http://arxiv.org/abs/1407.0945} {arXiv:1407.0945 [cond-mat.mes-hall]}
  \BibitemShut {NoStop}%
\bibitem [{\citenamefont {Fukaya}\ \emph
  {et~al.}(2016{\natexlab{a}})\citenamefont {Fukaya}, \citenamefont {Onogi},
  \citenamefont {Yamamoto},\ and\ \citenamefont {Yamamura}}]{Fukaya:2016dcl}%
  \BibitemOpen
  \bibfield  {author} {\bibinfo {author} {\bibfnamefont {H.}~\bibnamefont
  {Fukaya}}, \bibinfo {author} {\bibfnamefont {T.}~\bibnamefont {Onogi}},
  \bibinfo {author} {\bibfnamefont {S.}~\bibnamefont {Yamamoto}}, \ and\
  \bibinfo {author} {\bibfnamefont {R.}~\bibnamefont {Yamamura}},\ }\href
  {http://pos.sissa.it/cgi-bin/reader/conf.cgi?confid=256} {\bibfield
  {journal} {\bibinfo  {journal} {PoS}\ }\textbf {\bibinfo {volume}
  {LATTICE2016}},\ \bibinfo {pages} {330} (\bibinfo {year}
  {2016}{\natexlab{a}})},\ \Eprint {http://arxiv.org/abs/1612.00096}
  {arXiv:1612.00096 [hep-lat]} \BibitemShut {NoStop}%
\bibitem [{\citenamefont {Fukaya}\ \emph
  {et~al.}(2016{\natexlab{b}})\citenamefont {Fukaya}, \citenamefont {Onogi},
  \citenamefont {Yamamoto},\ and\ \citenamefont {Yamamura}}]{Fukaya:2016ofi}%
  \BibitemOpen
  \bibfield  {author} {\bibinfo {author} {\bibfnamefont {H.}~\bibnamefont
  {Fukaya}}, \bibinfo {author} {\bibfnamefont {T.}~\bibnamefont {Onogi}},
  \bibinfo {author} {\bibfnamefont {S.}~\bibnamefont {Yamamoto}}, \ and\
  \bibinfo {author} {\bibfnamefont {R.}~\bibnamefont {Yamamura}},\ }\href@noop
  {} {\  (\bibinfo {year} {2016}{\natexlab{b}})},\ \Eprint
  {http://arxiv.org/abs/1607.06174} {arXiv:1607.06174 [hep-th]} \BibitemShut
  {NoStop}%
\bibitem [{\citenamefont {Lian}\ and\ \citenamefont
  {Zhang}(2016)}]{PhysRevB.94.041105}%
  \BibitemOpen
  \bibfield  {author} {\bibinfo {author} {\bibfnamefont {B.}~\bibnamefont
  {Lian}}\ and\ \bibinfo {author} {\bibfnamefont {S.-C.}\ \bibnamefont
  {Zhang}},\ }\href {\doibase 10.1103/PhysRevB.94.041105} {\bibfield  {journal}
  {\bibinfo  {journal} {Phys. Rev.}\ }\textbf {\bibinfo {volume} {B94}},\
  \bibinfo {pages} {041105} (\bibinfo {year} {2016})},\ \Eprint
  {http://arxiv.org/abs/1604.07459} {arXiv:1604.07459 [cond-mat.mes-hall]}
  \BibitemShut {NoStop}%
\bibitem [{\citenamefont {Sen}\ and\ \citenamefont {Deb}(2012)}]{Sen:2012aa}%
  \BibitemOpen
  \bibfield  {author} {\bibinfo {author} {\bibfnamefont {D.}~\bibnamefont
  {Sen}}\ and\ \bibinfo {author} {\bibfnamefont {O.}~\bibnamefont {Deb}},\
  }\href {\doibase 10.1103/PhysRevB.85.245402} {\bibfield  {journal} {\bibinfo
  {journal} {Phys. Rev.}\ }\textbf {\bibinfo {volume} {B85}},\ \bibinfo {pages}
  {245402} (\bibinfo {year} {2012})},\ \Eprint {http://arxiv.org/abs/1203.3347}
  {arXiv:1203.3347 [cond-mat.mes-hall]} \BibitemShut {NoStop}%
\bibitem [{\citenamefont {Deb}\ \emph {et~al.}(2014)\citenamefont {Deb},
  \citenamefont {Soori},\ and\ \citenamefont {Sen}}]{Sen2014}%
  \BibitemOpen
  \bibfield  {author} {\bibinfo {author} {\bibfnamefont {O.}~\bibnamefont
  {Deb}}, \bibinfo {author} {\bibfnamefont {A.}~\bibnamefont {Soori}}, \ and\
  \bibinfo {author} {\bibfnamefont {D.}~\bibnamefont {Sen}},\ }\href {\doibase
  10.1088/0953-8984/26/31/315009} {\bibfield  {journal} {\bibinfo  {journal}
  {J. Phys.: Cond. Mat.}\ }\textbf {\bibinfo {volume} {26}},\ \bibinfo {pages}
  {315009} (\bibinfo {year} {2014})},\ \Eprint {http://arxiv.org/abs/1401.1027}
  {arXiv:1401.1027 [cond-mat.mes-hall]} \BibitemShut {NoStop}%
\bibitem [{\citenamefont {Wang}\ \emph {et~al.}(2014)\citenamefont {Wang},
  \citenamefont {Deng},\ and\ \citenamefont {Duan}}]{Wang:2014PRL}%
  \BibitemOpen
  \bibfield  {author} {\bibinfo {author} {\bibfnamefont {S.-T.}\ \bibnamefont
  {Wang}}, \bibinfo {author} {\bibfnamefont {D.-L.}\ \bibnamefont {Deng}}, \
  and\ \bibinfo {author} {\bibfnamefont {L.-M.}\ \bibnamefont {Duan}},\ }\href
  {\doibase 10.1103/PhysRevLett.113.033002} {\bibfield  {journal} {\bibinfo
  {journal} {Phys. Rev. Lett.}\ }\textbf {\bibinfo {volume} {113}},\ \bibinfo
  {pages} {033002} (\bibinfo {year} {2014})},\ \Eprint
  {http://arxiv.org/abs/1402.1204} {arXiv:1402.1204 [cond-mat.str-el]}
  \BibitemShut {NoStop}%
\bibitem [{\citenamefont {Qi}\ \emph {et~al.}(2008)\citenamefont {Qi},
  \citenamefont {Hughes},\ and\ \citenamefont {Zhang}}]{Qi:2008ew}%
  \BibitemOpen
  \bibfield  {author} {\bibinfo {author} {\bibfnamefont {X.-L.}\ \bibnamefont
  {Qi}}, \bibinfo {author} {\bibfnamefont {T.}~\bibnamefont {Hughes}}, \ and\
  \bibinfo {author} {\bibfnamefont {S.-C.}\ \bibnamefont {Zhang}},\ }\href
  {\doibase 10.1103/PhysRevB.78.195424} {\bibfield  {journal} {\bibinfo
  {journal} {Phys. Rev.}\ }\textbf {\bibinfo {volume} {B78}},\ \bibinfo {pages}
  {195424} (\bibinfo {year} {2008})},\ \Eprint {http://arxiv.org/abs/0802.3537}
  {arXiv:0802.3537 [cond-mat.mes-hall]} \BibitemShut {NoStop}%
\bibitem [{\citenamefont {Fu}\ and\ \citenamefont
  {Kane}(2008)}]{fu2008superconducting}%
  \BibitemOpen
  \bibfield  {author} {\bibinfo {author} {\bibfnamefont {L.}~\bibnamefont
  {Fu}}\ and\ \bibinfo {author} {\bibfnamefont {C.~L.}\ \bibnamefont {Kane}},\
  }\href {\doibase 10.1103/PhysRevLett.100.096407} {\bibfield  {journal}
  {\bibinfo  {journal} {Phys. Rev. Lett.}\ }\textbf {\bibinfo {volume} {100}},\
  \bibinfo {pages} {096407} (\bibinfo {year} {2008})},\ \Eprint
  {http://arxiv.org/abs/0707.1692} {arXiv:0707.1692 [cond-mat.mes-hall]}
  \BibitemShut {NoStop}%
\bibitem [{\citenamefont {Hashimoto}\ and\ \citenamefont
  {Kimura}(2016{\natexlab{a}})}]{Hashimoto:2016dtm}%
  \BibitemOpen
  \bibfield  {author} {\bibinfo {author} {\bibfnamefont {K.}~\bibnamefont
  {Hashimoto}}\ and\ \bibinfo {author} {\bibfnamefont {T.}~\bibnamefont
  {Kimura}},\ }\href {\doibase 10.1103/PhysRevB.93.195166} {\bibfield
  {journal} {\bibinfo  {journal} {Phys. Rev.}\ }\textbf {\bibinfo {volume}
  {B93}},\ \bibinfo {pages} {195166} (\bibinfo {year} {2016}{\natexlab{a}})},\
  \Eprint {http://arxiv.org/abs/1602.05577} {arXiv:1602.05577
  [cond-mat.mes-hall]} \BibitemShut {NoStop}%
\bibitem [{\citenamefont {Ryu}\ \emph {et~al.}(2010)\citenamefont {Ryu},
  \citenamefont {Schnyder}, \citenamefont {Furusaki},\ and\ \citenamefont
  {Ludwig}}]{Ryu:2010zza}%
  \BibitemOpen
  \bibfield  {author} {\bibinfo {author} {\bibfnamefont {S.}~\bibnamefont
  {Ryu}}, \bibinfo {author} {\bibfnamefont {A.~P.}\ \bibnamefont {Schnyder}},
  \bibinfo {author} {\bibfnamefont {A.}~\bibnamefont {Furusaki}}, \ and\
  \bibinfo {author} {\bibfnamefont {A.~W.~W.}\ \bibnamefont {Ludwig}},\ }\href
  {\doibase 10.1088/1367-2630/12/6/065010} {\bibfield  {journal} {\bibinfo
  {journal} {New J. Phys.}\ }\textbf {\bibinfo {volume} {12}},\ \bibinfo
  {pages} {065010} (\bibinfo {year} {2010})},\ \Eprint
  {http://arxiv.org/abs/0912.2157} {arXiv:0912.2157 [cond-mat.mes-hall]}
  \BibitemShut {NoStop}%
\bibitem [{\citenamefont {Atiyah}\ \emph {et~al.}(1964)\citenamefont {Atiyah},
  \citenamefont {Bott},\ and\ \citenamefont {Shapiro}}]{Atiyah:1964zz}%
  \BibitemOpen
  \bibfield  {author} {\bibinfo {author} {\bibfnamefont {M.~F.}\ \bibnamefont
  {Atiyah}}, \bibinfo {author} {\bibfnamefont {R.}~\bibnamefont {Bott}}, \ and\
  \bibinfo {author} {\bibfnamefont {A.}~\bibnamefont {Shapiro}},\ }\href
  {\doibase 10.1016/0040-9383(64)90003-5} {\bibfield  {journal} {\bibinfo
  {journal} {Topology}\ }\textbf {\bibinfo {volume} {3}},\ \bibinfo {pages}
  {S3} (\bibinfo {year} {1964})}\BibitemShut {NoStop}%
\bibitem [{\citenamefont {Ho\v{r}ava}(1999)}]{Horava:1998jy}%
  \BibitemOpen
  \bibfield  {author} {\bibinfo {author} {\bibfnamefont {P.}~\bibnamefont
  {Ho\v{r}ava}},\ }\href {\doibase 10.4310/ATMP.1998.v2.n6.a5} {\bibfield
  {journal} {\bibinfo  {journal} {Adv. Theor. Math. Phys.}\ }\textbf {\bibinfo
  {volume} {2}},\ \bibinfo {pages} {1373} (\bibinfo {year} {1999})},\ \Eprint
  {http://arxiv.org/abs/hep-th/9812135} {arXiv:hep-th/9812135 [hep-th]}
  \BibitemShut {NoStop}%
\bibitem [{\citenamefont {Kutasov}\ \emph {et~al.}(2000)\citenamefont
  {Kutasov}, \citenamefont {Marino},\ and\ \citenamefont
  {Moore}}]{Kutasov:2000aq}%
  \BibitemOpen
  \bibfield  {author} {\bibinfo {author} {\bibfnamefont {D.}~\bibnamefont
  {Kutasov}}, \bibinfo {author} {\bibfnamefont {M.}~\bibnamefont {Marino}}, \
  and\ \bibinfo {author} {\bibfnamefont {G.~W.}\ \bibnamefont {Moore}},\
  }\href@noop {} {\  (\bibinfo {year} {2000})},\ \Eprint
  {http://arxiv.org/abs/hep-th/0010108} {arXiv:hep-th/0010108 [hep-th]}
  \BibitemShut {NoStop}%
\bibitem [{\citenamefont {Hashimoto}\ and\ \citenamefont
  {Kimura}(2016{\natexlab{b}})}]{Hashimoto:2015dla}%
  \BibitemOpen
  \bibfield  {author} {\bibinfo {author} {\bibfnamefont {K.}~\bibnamefont
  {Hashimoto}}\ and\ \bibinfo {author} {\bibfnamefont {T.}~\bibnamefont
  {Kimura}},\ }\href {\doibase 10.1093/ptep/ptv181} {\bibfield  {journal}
  {\bibinfo  {journal} {PTEP}\ ,\ \bibinfo {pages} {013B04}} (\bibinfo {year}
  {2016}{\natexlab{b}})},\ \Eprint {http://arxiv.org/abs/1509.04676}
  {arXiv:1509.04676 [hep-th]} \BibitemShut {NoStop}%
\end{thebibliography}%


%

\end{document}